%% file: main.tex
\documentclass[sigconf, 10pt]{acmart}
\usepackage{tikz}
\usepackage{amsmath}
\usepackage{booktabs}
\usepackage[english]{babel}
\usepackage{blindtext}
\usepackage{pifont}
\usepackage{caption}
\usepackage{tcolorbox}
\usepackage{xcolor}
\usepackage{enumitem}
\usepackage{siunitx}
\usepackage{graphicx}
\usepackage{subcaption}
\usepackage{multirow}
\usepackage{multicol}
\usepackage{colortbl}
\usepackage{siunitx}
\usepackage{hyperref}
\usepackage{float}
\usepackage[ruled,vlined, linesnumbered]{algorithm2e}

\usepackage{algpseudocode}
\usepackage{makecell}
\usepackage{graphicx}
\usepackage{url}

\makeatletter
\pretocmd{\section}{\vspace{-0.2cm}}{}{}
\pretocmd{\subsection}{\vspace{-0.2cm}}{}{}
\pretocmd{\subsubsection}{\vspace{-0.3cm}}{}{}
\makeatother

\definecolor{LineNumberColor}{rgb}{0,0,1}

\newcommand{\systemname}{Jasper}

\renewcommand\footnotetextcopyrightpermission[1]{} 
\setcopyright{none}

\author{Haseeb Ashfaq}
\affiliation{
  \institution{NYU}
}
\thanks{\url{https://haseeblums.github.io}}

\author{Jinkun Geng}
\affiliation{
  \institution{Stanford University}
}

\author{Daniel Duclos-Cavalcanti}
\affiliation{
  \institution{Technical University of Munich}
}

\author{Ulysses Butler}
\affiliation{
  \institution{NYU}
}

\author{Xiyu Hao}
\affiliation{
  \institution{NYU}
}

\author{Radhika Mittal}
\affiliation{
    \institution{UIUC}
}

\author{Srinivas Narayana}
\affiliation{
  \institution{Rutgers University}
}

\author{Anirudh Sivaraman}
\affiliation{
  \institution{NYU}
}

\begin{document}

\let\oldcite\cite
\renewcommand{\cite}[1]{\textcolor{darkred}{\oldcite{#1}}}

\let\oldref\ref
\renewcommand{\ref}[1]{\textcolor{darkgreen}{\oldref{#1}}}

\newcommand{\todo}[1]{{\textcolor{red}{TODO: #1}}}

\newcommand{\mh}[1]{
{\textcolor{brown}{MH: #1}}
}

\newcommand{\radhika}[1]{
{\textcolor{magenta}{RM: #1}}
}

\definecolor{darkgreen}{rgb}{0.0, 0.5, 0.0}
\definecolor{darkred}{rgb}{0.6, 0.0, 0.0}
\algrenewcommand{\alglinenumber}[1]{\color{red}\footnotesize#1:}

\title{Design And Implementation Of A Scalable Financial Exchange In The Public Cloud}

\newenvironment{parafont}{\fontfamily{ptm}\selectfont}{}
\newcommand{\Para}[1]{\vspace{2pt}\noindent\begin{parafont}\textbf{\textit{#1}}\end{parafont}}
\newcommand{\sysname}[0]{Jasper\xspace}

\begin{abstract}
    \input{abstract}
\end{abstract}

\maketitle
\pagestyle{plain}

\input{intro}
\input{background}
\input{system_design}
\input{multicast_tree}
\input{dyn_relationships}
\input{vm_hedging}
\input{receiver_hedging}
\input{multicast_packet_losses}
\input{new_orders_submission}
\input{evaluations}
\input{related_work}
\input{conclusion}

\bibliography{main}
\input{appendix}

\end{document}

%% file: abstract.tex
We present \sysname{}, a system for meeting the networking requirements of financial exchanges on the public cloud. \sysname{} uses an overlay tree to multicast market data from an exchange to 1000 participants with $\leq\SI{1}{\micro\second}$ difference in data reception time between any two participants, crucial for maintaining fair competition. \sysname reuses the same tree for scalable inbound communication (participants to exchange), introducing a scheduling policy to enhance an exchange's throughput during periods of bursty traffic.  It also presents a message sequencing mechanism that achieves globally ordered delivery of messages from participants to the exchange, i.e., messages of participants are seen by the exchange in order they are generated to maintain fairness. 
\sysname{} achieves better scalability and $\approx$50\% lower latency than the AWS multicast service~\cite{aws_tg_multicast}. \sysname{} outperforms existing system, CloudEx~\cite{cloudex-hotos} in terms of supported number of participants, order matching rate of the exchange and multicast latency. \sysname{}'s techniques can be applied to other existing systems (e.g., DBO) to enhance their performance. 

%% file: intro.tex
\section{Introduction}


Financial exchanges are migrating to the public cloud for reasons such as improved scalability and reduced capital expenditure. Despite its benefits, the public cloud poses unique challenges. Exchanges have traditionally operated in on-premise or colocation facilities, engineered for deterministic and low latency. For fair market access, exchanges equalize cable lengths between the exchange and participant servers. This approach ensures that (i) all participants receive market data from the exchange simultaneously (\textit{outbound fairness}) and (ii) an order generated earlier by one participant reaches the exchange before orders generated later by other participants (\textit{inbound fairness}). It is common to employ low jitter switches (e.g., L1 switch~\cite{l1_switch}) to reduce latency variance. 

\begin{figure}[t]
\centering
\includegraphics[width=0.45\textwidth]{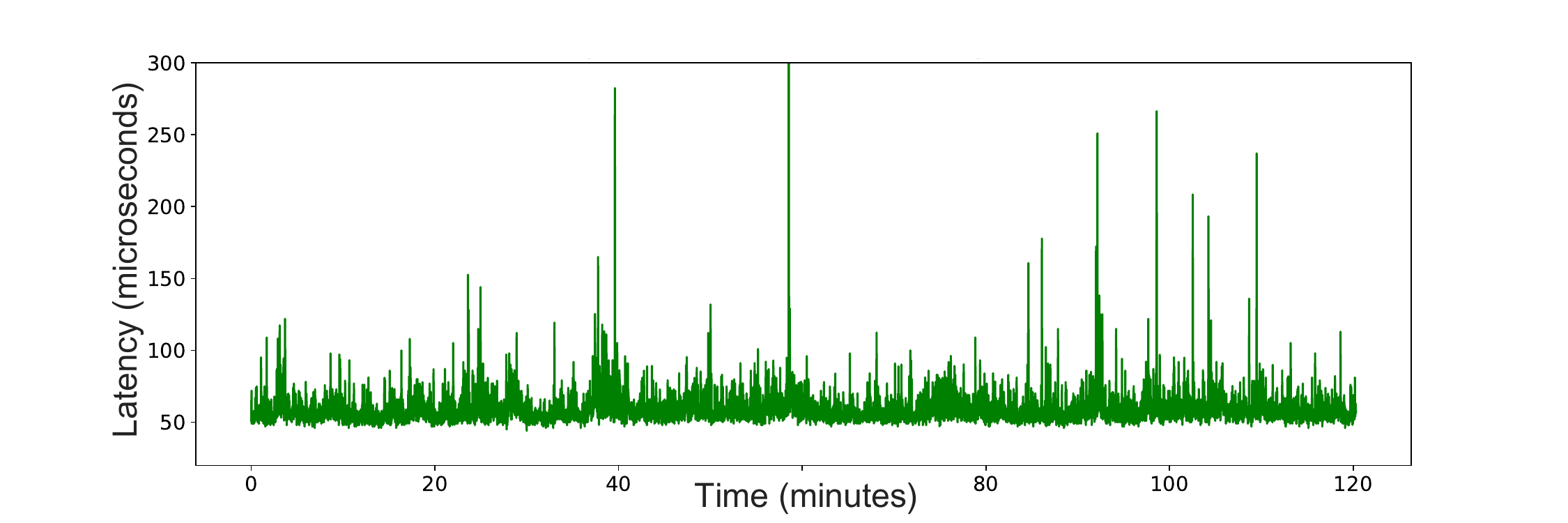}
\vspace{-0.3cm}
\caption{Latency between a pair of VMs varies over time}
\label{fig:latency_aws}
\vspace{-0.5cm}
\end{figure}

However, the public cloud lacks these enhancements. It is a best-effort environment characterized by nondeterminism (e.g. latency variance in Figure~\ref{fig:latency_aws}). In response, several projects have developed techniques for cloud-based exchanges. These include using synchronized clocks to compensate for nondeterminism~\cite{cloudex-hotos}, using SmartNICs to hold data until all receiver has received it~\cite{octopus_google}, and new fairness definitions~\cite{dbo_hotnets, dbo_prateesh}. 
These projects have demonstrated promising results for tens of participants but exhibit significant performance degradation as the number of participants increases. This limitation arises because scalability was not a primary design objective. Instead, the initial focus was on establishing a functional proof of concept for a fair exchange on the cloud. Having made significant progress in that regard, the next logical step is to consider scalability.
Consequently, our paper develops techniques for \emph{scale}: \emph{How do we architect an exchange to support communication between the exchange and \textasciitilde1000 participants in the cloud while ensuring fairness in the exchange?} In practice, a scalable exchange would also need scale-out \emph{compute} techniques for the exchange server---a concern that is out of scope.

Our system, \sysname{}, tackles two major challenges to provide network support for scalable exchanges. First, how do we support \emph{outbound} communication of market data from the exchange to 1000 market participants, while ensuring (a) low spatial variance, i.e., all participants receive market data nearly simultaneously, (b) low latency from the exchange server to the participant, and (c) low temporal variance, i.e., latency doesn't fluctuate over time? Second, how do we support \emph{inbound} communication of participant orders to the exchange while (a) providing a fair chance of making trades to all the traders and, (b) achieving high throughput for the exchange server ---especially during intense market activity when bursts of orders from many participants arrive at once, causing incast-style~\cite{incast} drops and increased latency? \sysname{} integrates multiple well-established mechanisms within a novel context as well as introduces new techniques to achieve high performance for a cloud-hosted exchange.


First, to scale outbound communication from the exchange to a large number of participants, we employ an overlay multicast tree composed of a root exchange VM, proxy VMs as intermediate nodes, and participant VMs as leaf nodes. We develop a simple and effective heuristic to tune the tree's fan-out and depth, navigating a trade-off between increased serialization delay due to greater fan-out and increased propagation delay because of greater depth. To lower latency and counter the cloud's variability, we pervasively employ \emph{hedging}: routing redundant copies of the market data through multiple proxy and receiver VMs and rotating parent-to-child associations at each tree level on each multicasted packet.

Second, to provide scalable and fair inbound communication from participants to the exchange, we propose a sequencer that relies on recent advancements in clock synchronization that is robust to latency fluctuations and works for VMs without hardware support. The sequencer ensures that an exchange server sees the messages generated by participants in their generation order so that \textit{inbound fairness} is achieved regardless of the arbitrary message delays. We also propose a scheduling policy, Limit Order Queue, that helps achieve fairness and high order matching rate of the exchange during periods of bursty market activity. Finally, we reuse the overlay multicast tree in the \emph{reverse direction} to proxy participants messages to the exchange which helps against incast-style packet drops by reducing the fan-in at the exchange when supporting a large number of participants. This leads to high throughput for the exchange server and low latency for trade orders. 



\sysname{} can support a maximum throughput of 175K multicast messages per second at which point it is limited by a proxy VM's egress bandwidth. It also scales to 1000 receivers/VMs, achieving a median multicast latency of $\leq\SI{250}{\micro\second}$ while maintaining a latency difference of $\leq\SI{1}{\micro\second}$ across these receivers. \sysname{} also scales better and achieves 50\% lower latency compared to AWS' Transit Gateway-based multicast. On the inbound side, \sysname{} efficiently handles large bursts of orders, using LOQ to increase the order matching rate proportional to the bursts and maintain low order latencies. \sysname{} outperforms a prior system, CloudEx~\cite{cloudex-hotos} in terms of scalability, order matching rate and multicast latency. More importantly, the techniques described in \sysname{} are meant to provide a networking layer for the exchange systems and are thus composable with the existing exchange systems to enhance their performance. We think of \sysname{} as CloudEx augmented with several scaling methodologies that are able to maintain fairness. Onyx will be opensourced.

%% file: background.tex

\section{Background}
\Para{Financial Exchange Setup.} An exchange typically has an exchange server and multiple market participant (\textbf{\textit{MP}}) servers. The exchange server runs a matching engine (\textbf{\textit{ME}}) to process trading orders from the MPs and multicast market data to them. Market data reveals market state e.g., asset prices and processed orders. Orders can be bid orders, which aim to purchase an asset at a specific price, and ask orders, which aim to sell an asset at a specific price.

The ME maintains a limit order book (\textbf{\textit{LOB}}) (Fig.~\ref{fig:lob}), which lists all bid and ask orders from MPs. When a bid price exceeds or match the lowest ask price, the orders are executed (and matched together). Unexecuted orders remain in the LOB, waiting for a match. In the price-time-priority algorithm~\cite{price_time_priority},\footnote{Price-time-priority is widespread, so we use it. Alternatives exist such as the pro-rata algorithm~\cite{price_time_priority}} orders are arranged in price levels, with those at the same level sorted by their arrival time at the exchange (which is equivalent to sorting them by their generation time at the MPs if all cables connecting MPs to exchange are precisely equalized). An LOB snapshot (Fig.~\ref{fig:lob}) shows a separation between bid and ask price levels, with the mid-price indicating the asset's true value. Thus, orders closer to the mid-price have higher chances of getting matched early. Any scheduling policy on orders should ensure that the semantics of price-time priority matching algorithm are not impacted i.e., the sequence of matched orders should remain unchanged. 

Traditionally, the exchange and MP servers are co-located in the same datacenter and connected via equal-length wires using L1~\cite{l1_switch} switches with switch multicast. The overall infrastructure is designed to provide low and equal latency from the exchange server to all MPs in both directions. 

\Para{Challenges of Cloud Migration.} While the public cloud offers many advantages, it does not offer low-level control, e.g., allowing tenants to control wire lengths. The public cloud also exhibits high latency variance~\cite{cloudy_forcast} and computation variance~\cite{nsdi16_ernest}. The latency between a pair of VMs can be significantly different from another pair of VMs~\cite{vig_thesis}. Latency also fluctuates over time: Figure \ref{fig:latency_aws} plots the $90p$ latency between a pair of VMs in an AWS region for a tumbling window of \SI{1}{\second}. The figure also shows infrequent but unpredictable latency spikes that substantially increase latency~\cite{cloudy_forcast, dbo_prateesh}. Such phenomena make it challenging to achieve low and deterministic latency in the public cloud. 

Given the above challenges, it is reasonable to ask whether financial exchanges should ever be migrated to the cloud --- and if so, how. This is an ongoing debate with reasonable arguments on both sides. Beyond the research projects~\cite{cloudex-hotos, dbo_prateesh, octopus_google} discussed in the following, some cloud providers are exploring close partnerships to build private clouds for exchanges' bespoke requirements~\cite{cme_google_colab, google-invest-financial-exchange}. \systemname{} contributes to this debate by studying what performance guarantees can be achieved if \emph{a cloud tenant designs the exchange architecture on the public cloud with publicly available cloud APIs}. With \sysname{}'s DIY approach, a tenant does not need to wait for a cloud provider to build private clusters and can quickly ramp up a performant and fair exchange. 

\Para{Is cloud migration still relevant?} There has also been a debate on cloud migration and cost cuts being a paradox, giving emergence to \emph{cloud repatriation} (moving away from cloud) as discussed by Martin Casado~\cite{martin_casado_cloud_repatriation}. Cloud repatriation, unlike cloud migration, is highly relevant to large SaaS companies like Snowflake or Dropbox, which operate at a scale where they can benefit from economies of scale by owning infrastructure and have consistent workloads that justify such investments. However, this narrative doesn’t extend as well to smaller entities like financial institutions, which lack the consistent, cloud-scale workloads to make private infrastructure cost-effective and often benefit more from the flexibility and cost savings of not managing their own hardware.

\Para{Prior Work and \sysname's Motivation.} 
CloudEx~\cite{cloudex-hotos}, the first cloud-hosted exchange system, may lose inbound as well as outbound fairness under latency fluctuations. To maintain fairness, participants wait until a set timeout (to counter latency fluctuations) before processing any order to ensure every participant has received the market data. The matching engine waits until a set timeout before processing any order to ensure that all the earlier generated orders have been received. As the number of participants increase, these timeouts need to increase (to not lose fairness) accordingly which lead to high latency and low order matching rate. 
DBO~\cite{dbo_prateesh} leverages mechanisms to always guarantee fairness among participants by decoupling fairness from latency fluctuations, although it is only applicable to a subset of trades that depend on the last received market data batch. As the number of participants increase, DBO also suffer from degraded performance because of (i) incast on the inbound side and, (ii) large latency of market data. Both of these systems can benefit from techniques to enhance the number of participants that maintain low latency, low latency fluctuations and high throughput for the exchange. 

Consequently, \sysname's main contribution is architecting for scale while maintaining fairness and achieving high performance. \sysname adopts the idea of synchronized clocks from CloudEx, but scales the system much further by using a communication tree in both the inbound and outbound directions to achieve high performance with a large number of participants. \sysname{} augments the tree with (i) a new sequencer to ensure inbound fairness under latency fluctuations (ii) a scheduling algorithm to achieve high performance under bursts of orders and, (iii) several performance variance mitigation techniques to help achieve fairness. 

%% file: system_design.tex

\begin{figure*}[!t]
    \centering
    \begin{minipage}{1\textwidth}
        \centering
        \includegraphics[width=1\linewidth]{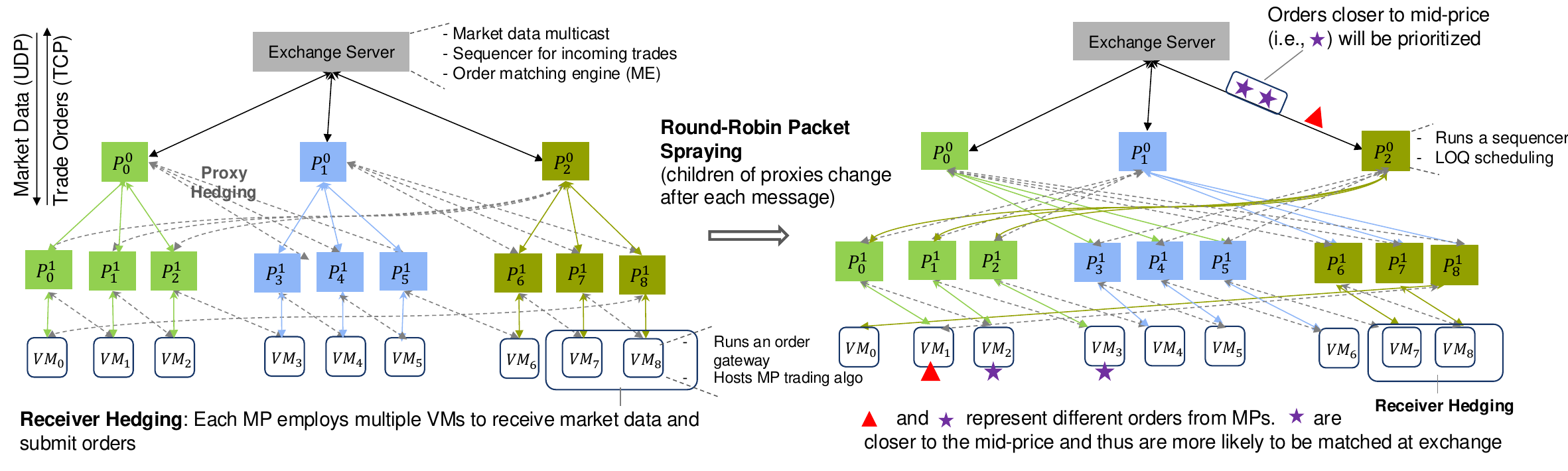}
        \vspace{-0.6cm}
        \caption{Overview of \systemname{}.
        }
        \label{fig:tree_overview}
    \end{minipage}%
\end{figure*}

\section{\sysname Overview}
\label{sec:overview}

\Para{Setup and goals.} Each MP VM hosts an order gateway and a trading algorithm.\footnote{``trading algorithm'' and ``MP'' are interchangeably used} Clocks of all MP VMs and the exchange server are synchronized using Huygens algorithm~\cite{huygens},  which enables nanoseconds level synchronization for VMs without hardware support. Trading algorithm generates the orders and submit it to the colocated gateway. A gateway then attaches its current timestamp to an order and forwards it to the exchange (via TCP~\cite{trading_hotnets}), which hosts the matching engine (ME) to process orders. MP VMs are controlled by the exchange which loads the trading algorithms in them. Market data from the exchange (via UDP~\cite{trading_hotnets}) first arrives at a gateway which then forwards it to the trading algorithm. Such a model has been proposed previously~\cite{jasper_poster, yu2024cuttlefish}.\footnote{An avid reader may be interested in a brief discussion of several deployment/trust models and their performance in Appendix~\ref{sec:trust_model}.} \sysname aims to provide fairness during the competition among MPs for both the outbound (exchange to MPs) and inbound (MPs to exchange) directions. We specify the inbound/outbound fairness as follows. 

\begin{definition}[Outbound Fairness]
Every market data message sent from the exchange server to the market participants (MPs) should be seen by all the MPs simultaneously. 
\end{definition}
\begin{definition}[Inbound Fairness]
An order generated earlier than other orders should be processed by the matching engine earlier than the other orders, irrespective of which MP generated which order. 
\end{definition}

While respecting the above fairness definitions, \sysname aims to achieve high performance i.e., supporting large number of MPs and achieving a high order matching rate, setting it apart, but complementary, from the previous work. Figure \ref{fig:tree_overview} presents an overview of \systemname{} architecture. \systemname{} employs a bidirectional overlay tree: the exchange server is the root and MPs are the leaves, with intermediate proxy nodes. It is well known that trees help scale communication to many receivers~\cite{delay_multicast_mesh_overlays, delay_and_delay_variation_multicast}. We build on such a tree to provide a multicast service for market data and handle order submissions from a large number of MPs, but adapt the tree to the high-variance environment of the public cloud. 

\Para{Outbound: ME to Participants (market data multicast).} On the outbound side of an exchange, we augment the base tree with 3 techniques to lower multicast latency and minimize the latency variance: (i) round-robin packet spraying, (ii) proxy hedging and (iii) receiver hedging. All 3 hedge the risk of some part of the system exhibiting performance variance.

To minimize the impact of latency spikes on the links between the tree nodes, proxies at each layer rotate their child VMs after each message, ensuring messages do not always follow fixed paths. This approach limits the effect of latency spikes, as only a small subset of messages is impacted by any transient spike on any link within the tree.

We employ proxy hedging to mitigate the impact of straggler proxies that may introduce latency inflation. In this approach, each VM (a child proxy or a receiver) gets copies of a message from multiple proxies. Each VM processes the first received copy, discards the others, and forwards this copy to its children and some nieces. This way, if a proxy inflates latency for all messages going through it, its children can still receive market data promptly from an aunt proxy.

Lastly, if an MP's VM becomes a straggler, it will not be able to compete effectively because of increased latency for its market data. We hedge this risk by allocating two VMs to each MP, where both run the same trading algorithm. 

\Para{Inbound: Participants to ME (orders submission).} We develop a sequencer that sits at the ingress of the exchange server. All the incoming orders from MPs are fed to the sequencer while the output (the sequenced orders) are processed by the matching engine (ME) hosted by the exchange server. The sequencer ensures that inbound fairness is achieved regardless of latency fluctuations in the paths of orders. During periods of bursty market activity, MPs may generate a lot of orders overwhelming the ingress of the exchange. As we employ TCP in the inbound communication, the overwhelming of the exchange leads to queue build ups at the MP VMs. A special priority queue, Limit Order Queue (LOQ), runs at the egress of each MP VM for servicing the built up queues so that latency of orders remains low and high order matching rate at ME is achieved. 

LOQ distinguishes critical orders --orders that will get matched by the ME upon their arrival at the exchange server-- from non-critical orders-- orders that will just stay in the limit order book and will not get immediately matched. LOQ prioritizes critical orders over noncritical orders so that ME processes the matchable orders earlier than non-matchable ones, leading to high order matching rate and low latency for orders. As we explain later, LOQ scheduling is designed to not affect inbound fairness. 

Further, as the number of MPs increases, the ingress of the exchange server becomes a prominent bottleneck because of the large number of MPs simultaneously submitting orders. Even if the cumulative load of all the MPs is well under the ingress capacity of the exchange, the pattern of large number of orders from a large number of MPs simultaneously arriving at the exchange leads to incast-style packet drops. To resolve this, we reuse the proxy tree but in the reverse direction (for MP-to-exchange communication): MPs submit their orders to their parent proxies where it travels up the tree and reaches the ME at the root. Using a tree in the reverse direction reduces the fan-in factor of ME: the ME has to receive and process streams of orders from a small number of proxies instead of all MPs. This reduces packet losses, leads to fewer retransmissions by TCP which sums up to high throughput for the exchange server. 

With the tree in place, we see queue build ups at the proxy nodes during bursty order arrivals. We apply our above-mentioned priority queue, LOQ, at the proxy nodes as well where it services the queues to significantly improve the order matching rate of the ME.


%% file: multicast_tree.tex
\section{Market Data Multicast}
\label{sec:tree_low_latency}
\label{sec:exchange_outbound}

To satisfy outbound fairness, market data should be distributed \emph{simultaneously} to all MPs. We develop overlay techniques atop a best-effort fabric to achieve fair and low-latency multicast that scales to a large number of receivers. 


Due to the lack of switch support, the multicast in the cloud can only be implemented by using multiple direct unicasts. Since the back-to-back unicasts are serialized over the sender's egress, the latter receivers (among a large number of receivers) will receive the unicast message much later than the others due to the cumulative serialization delay at the sender. To tackle this unfairness issue, we choose to use an overlay tree to implement the scalable multicast. As illustrated in Figure~\ref{fig:tree_overview}, the sender sits at the root of the tree and only sends its messages to a limited number of proxies. Each proxy then relays the message down the tree to the lower-layer proxies, and recursively down to receiver VMs at the leaves. Since each node's fanout is limited, the cumulative serialization delay is constrained at each layer, reducing the heterogeneity of message delays among receivers. 





\begin{figure}[!t]
    \centering
    \begin{minipage}{0.25\textwidth}
        \centering
        \includegraphics[width=\textwidth]{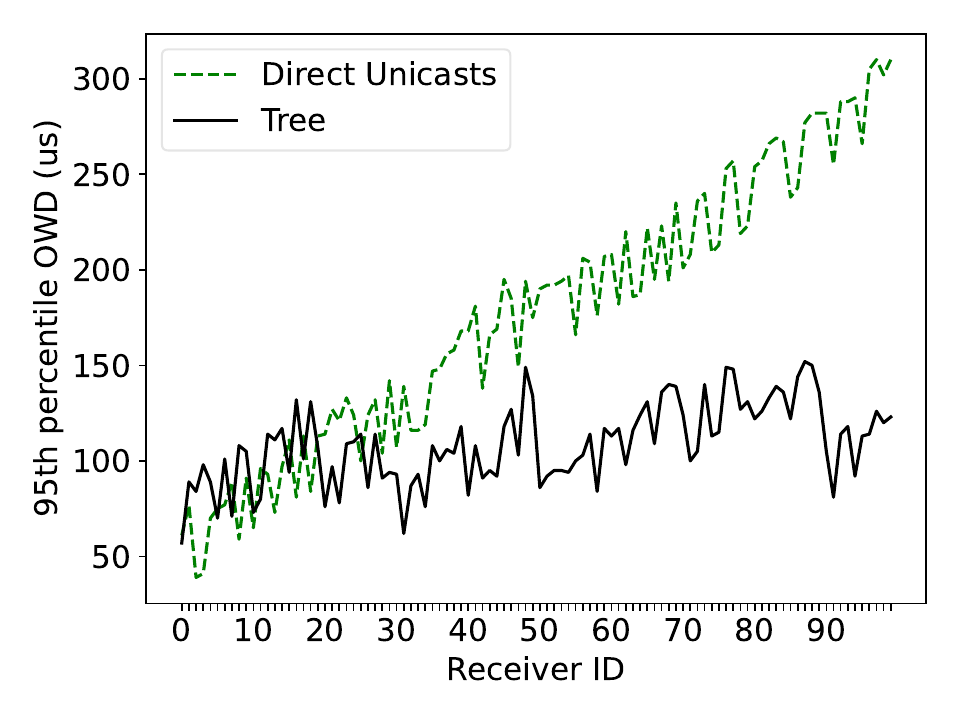}
        \vspace{-0.6cm}
        \caption{Direct unicasts do not scale well; a tree does. }
        \label{fig:socket-spray}
    \end{minipage}%
    \hfill
    \begin{minipage}{0.2\textwidth}
        \centering
        \includegraphics[width=\linewidth]{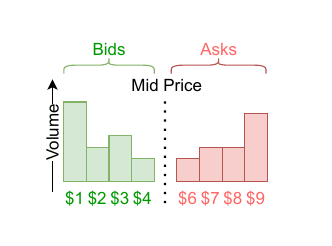}
        \caption{A Limit Order Book snapshot.}
        \label{fig:lob}
    \end{minipage}
    \vspace{-0.5cm}
\end{figure}

Tuning the tree's fan-out ($F$) and depth ($D$) is crucial for minimizing end-to-end multicast latency for a given number of receivers ($N$) as extra hops in the path of messages increase latency. We conducted experiments with various <$D, F$> for $N=10, 100$, and $1000$ and verified that $F \approx 10$ and $D = \left\lceil \log_{10} N \right\rceil$ yield near-minimum latency, and more sophisticated strategies bring little gains beyond that (Appendix~\ref{app:decide_d_and_f}) because of the cloud's inherent variability. 

We synchronize the clocks of all MPs, and the ME using Huygens algorithm~\cite{huygens}. We follow CloudEx's ``hold-and-release'' protocol~\cite{cloudex-hotos} and assign each multicast message a future time\-stamp called a deadline. After receiving a message, the gateways hold it until its deadline before forwarding to the MP's trading algorithm.  The deadline is calculated by adding a \emph{headroom} to the sending time of the market data. To ensure all receivers likely receive a message before the deadline, the headroom is decided by the maximum latency from the sender to every receiver. 

However, since \sysname targets a much larger number (\textasciitilde1000) of receivers than CloudEx(\textasciitilde10), we realize that simply applying the ``hold-and-release'' is not sufficient. As the number of receivers increases and the latency exhibits high variance, the required holding duration by a receiver increases to ensure simultaneous delivery to all receivers, which inflates the multicast latency. Tree helps in lowering the end-to-end latency, however latency variance still needs to be minimized. So, we incorporate three optimizations into our proxy tree to reduce both the end-to-end latency and the latency variance. 



%% file: dyn_relationships.tex
\subsection{Round-Robin Packet Spraying}
\label{sec:dyn-relationships}
\label{rrps}


In order to reduce latency spikes and alleviate the impact of bursts of market data, we develop round-robin packet spraying (RRPS). In RRPS, parent-child links in the tree are flexible, i.e., proxies change their children after each multicast message. Specifically, for a given proxy, on every new multicast message, the set of children is circularly shifted by 1. For example, the first proxy in a layer will have the first $F$ proxies of the next layer as its children for message 0, the next $F$ proxies as children for message 1, etc. This ensures that each proxy has a parent for each message, while the set of children for a parent proxy continuously changes. With RRPS, a single leaf node in a proxy tree has a total of $\prod_{d=1}^{D-1} F^d$ different virtual paths starting from the root node. Successive messages traverse this abundance of virtual paths in a round-robin fashion. By contrast, without RRPS, there is only one virtual path from the root to each leaf node and all messages traverse that path. The increased path diversity provides two major benefits for latency reduction. 

First, if a spike occurs on any VM-to-VM virtual link, only a subset of messages is impacted: Since messages are round-robined across all available virtual paths, many messages will take alternative paths that avoid this virtual link.\footnote{Here, we assume that spikes are characteristics of individual virtual links, not VMs themselves---as is often the case. If spikes were a VM characteristic---as is also the case sometimes---many virtual links would simultaneously be affected, reducing the benefits of RRPS. For such \emph{straggler} VMs, we use proxy hedging (\S\ref{sec:hedging-design}).}

Second, RRPS uncovers and utilizes new virtual paths that were unused earlier. For example, consider two adjacent proxy layers: a parent layer with $P$ proxies and a children layer with $P \times F$ proxies where each parent proxy has $F$ children. Without RRPS, the number of virtual paths utilized by messages going from parents to children is $P\times F$. With RRPS, there are $P\times P \times F$ virtual paths. This allows us to distribute bursts among many more virtual paths, which reduces queue build-up and latency. 

At its core, RRPS works because there is an abundance of virtual paths available within an overlay proxy tree. This is because every VM in a cloud region can communicate with every other VM since a VM-to-VM link is virtual---effectively the network topology is a clique. This is unlike the links in physical networks~\cite{best_paths_harry} where the network topology is not a clique and is limited by physical constraints like geographic distance and cables.

%% file: vm_hedging.tex
\subsection{Proxy Hedging}
\label{sec:vm_hedging}
\label{sec:hedging-design}
\label{proxy_hedging}

A VM's performance in the cloud can degrade due to factors like noisy neighbors or live VM migration, causing latency fluctuations for all messages passing through the VM~\cite{lemondrop}. To mitigate this, we develop proxy hedging, where each node in the tree receives multiple copies of a message from proxies in the higher layer. A node processes only the first copy and discards duplicates, thus reducing the impact of any straggler proxy VMs. It decreases both temporal and spatial latency variance, leading to more stable latency and reducing the delivery window size (i.e., the difference between the earliest and latest receiver's latency). For ease of explanation, we assume RRPS is not in effect, but the 2 techniques compose.

In proxy hedging, each proxy node sends messages to the children of $H$ of its siblings along with sending messages to children of its own where $H$ is the hedging factor. For example, in Figure~\ref{fig:tree_overview} (left),  when hedging is not enabled (i.e., $H=0$), proxy $P_3^{1}$ ($3^{rd}$ proxy in $1^{st}$ layer) only receives messages from $P_1^{0}$. As a result, $P_3^{1}$ may suffer from high latency if $P_1^{0}$ or the path from $P_1^{0}$ to $P_3^{1}$ encounters latency fluctuations. By using hedging, $P_3^{1}$ not only receives messages from $P_1^{0}$, but also receives the same messages from one (if $H=1$) other node, $P_0^{0}$. This technique shows significant benefits that we evaluate in \S\ref{evals_hedging}. 

A proxy processes the earliest received copy among $H+1$ copies of a message and discards the rest, achieving significant latency reduction of each message, at the cost of some throughput wastage. We find out empirically as well as with Monte Carlo simulation that a small $H$ (of up to 2) is sufficient for reaping its most benefits. A Monte Carlo analysis of hedging is presented in Appendix~\ref{app:hedging_analysis} that corroborates: (i) latency and its variance reduces as $H$ increases, and (ii) increasing $H$ shows diminishing returns. 

Unlike the previous technique, RRPS, proxy hedging leads to redundant work which decreases the effective throughput (i.e., goodput) of a proxy. The goodput of a proxy node can be defined as: $\frac{1}{H + 1} \times \emph{throughput}$ because each proxy sends redundant messages to the children of $H$ siblings. One way to recover the lost throughput is to employ several parallel trees that share the root and the leaves (assuming the leaves have enough ingress bandwidth). The increased number of proxies also raises concerns of dollar cost, which could potentially be a drawback here. Appendix~\ref{app:cost_hedging} includes the detailed estimation of monetary cost.

%% file: receiver_hedging.tex
\subsection{Receiver Hedging}
\label{sec:receiver_hedging}

If a VM belonging to an MP becomes a straggler, the MP would become more likely to lose the trading competition against the other MPs, because the market data it receives will lag behind other MPs. To hedge this, we provide each MP with two receiver VMs; similar measures are taken in on-premises exchanges so we do not expand more on it. This simple technique enhances performance significantly at the tail, more details in Appendix~\ref{app:receiver_hedging_sync}.

%% file: multicast_packet_losses.tex
\subsection{Remarks on Multicast Packet Losses}
\label{sec:packet_loss_multicast}

In on-premises exchanges, the multicast service does not provide reliable delivery and \sysname{} adopts the same (and uses UDP in the outbound communication). Instead, a separate "rewinder" service is deployed that handles any retransmission requests from MPs~\cite{rewinder_service, brian_nigito_interview}. MPs use sequence numbers of the messages to detect gaps and request a retransmission. The exchange infrastructure is engineered to provide negligible packet losses so the MPs only have to contact the rewinder rarely. \sysname{} multicast service needs to similarly provide low losses. Our hedging techniques help lower the packet losses, but we observe that even without our hedging, the losses in outbound direction are small: we observe 0.007\% lost packets in a \SI{20}{\second} benchmark on GCP while exhausting 80\% of egress bandwidth (details in Appendix~\ref{app:packet_losses}). 

%% file: new_orders_submission.tex
\section{Orders Submission Service}

On the inbound side, we aim to provide the order submission service that can achieve both inbound fairness and efficient order matching. To do so, we introduce a sequencer at the exchange server to enable fair processing of orders. Meanwhile, we install a novel scheduling policy, namely Limit Order Queue (LOQ) policy, on each gateway, which can significantly improve the ordering matching rate of the exchange. For simplicity, we will first describe the sequencer and LOQ in a setting where gateways are directly connected to the exchange server. Then, we continue to explain how the multicast tree is integrated to substantially improve the performance of the order submission service.


\subsection{Sequencer}
\label{sequencer}
The exchange system has to provide a fair chance of trading to all the MPs, which requires the exchange to process MPs' orders following their generation order. To satisfy this inbound fairness requirement, on-prem exchanges equalize the latency between MP servers and the exchange server by connecting them with the cables of the same length. However, such an approach is not feasible in the public cloud where the tenants have no access to the underlying infrastructure. Therefore, we seek an alternative approach to achieving inbound fairness by using synchronized clocks, which have become available in today's public cloud.

In \sysname, we synchronize the clocks on both MPs and the exchange VMs with an accuracy of 10s of nanoseconds using Huygens~\cite{huygens},\footnote{Given that we deal with latencies on the order of microseconds, such a clock synchronization accuracy is exceptionally good.} a network effect based clock synchronization algorithm robust to latency variance. We place a sequencer at the exchange server, which aims to hold the incoming orders and releases the orders according to the global FIFO order i.e., the sequencer ensures that an exchange sees the orders with timestamps in the non-decreasing order. The sequencer only releases an order of an MP to the ME when (i) it sees higher timestamped orders from every other MP and the (ii) the current order has the lowest timestamp among the orders present at the sequencer.~\footnote{ The ties on the timestamps are broken based on the MP ID.} For abetting liveness (i.e., sequencer does not block processing of some orders for long periods), gateways periodically generate dummy orders on behalf of inactive traders -- similar to the heartbeats employed by DBO~\cite{dbo_prateesh}. Given synchronized clocks and ordered delivery per participant (e.g., via TCP), the sequencer ensures \emph{safety}, i.e., inbound fairness. This is different from the prior works (\S\ref{related_work}): (i) CloudEx violates safety even with accurate clock synchronization because it waits for orders only until a set timeout and, (ii) DBO holds safety for only a subset of trades (i.e., trades that only depend on specific market data points), although it does not assume synchronized clocks. \sysname's safety comes at the cost of liveness as a failure of a gateway will block the sequencer. As the common practice in distributed systems~\cite{podc09-vertical-paxos, nsdi23-hydra, cmak,zookeeper}, the failure detection of gateways is be conducted by a standalone \emph{configuration manager} (e.g., Zookeeper), which periodically checks the health of each gateway and reconfigures the membership for the \sysname cluster. We omit the details of the configuration manager since it is beyond the scope of this work. The sequencer protocol runs as follows. 


\input{sequencer_protocols}

\subsubsection{\textbf{Implementation}}
Algorithm \ref{alg:sequencer} presents an efficient implementation of the sequencer's enqueue and dequeue routines. \textbf{Enqueue} is made lightweight so that incoming messages can be processed as quickly as possible and a queue formation at the ingress of the exchange's VM can be avoided. \textbf{Enqueue} is invoked at each new message while \textbf{Dequeue} runs in a separate thread indefinitely. 



\subsection{Limit Order Queue}
\label{limit_order_queue}

As the common practice in today's exchange systems, we employ TCP to provide reliable delivery of orders (from MP to exchange). During periods of bursty activity in the market, incast congestion at the exchange server's ingress occurs, leading to retransmissions which in turn leads to queue build-ups at the gateways (because of backpressure). To tackle the queuing delays incurred at the gateways, we develop a scheduling scheme --Limit Order Queue (LOQ) scheduling--to schedule the orders at each gateway, which can effectively reduce order matching latency and improve the order matching rate at ME during periods of bursts. 



\input{loq_protocol}
\subsubsection{\textbf{Inbound fairness under LOQ}}
For safety, sequencer assumes that messages of a single MP arrive at the sequencer in their generation order. As LOQ at a gateway reorders the messages, this assumption is violated. However, we claim that fairness is still achieved as we have designed the LOQ policy to do so. 
Here we provide the intuition of our claim while Appendix~\ref{loq_proof_intuition} contains a detailed intuitive proof. 

Assume there is a static mid-price. Instead of looking at the order in which messages/orders are presented to the ME, let's look at the order in which the orders are executed by the ME. By design LOQ ensures that if an order is selected for execution by ME, then all the older executable orders must have been executed already. Executable orders are the ones whose price falls in [$m-w$, $m+w$] i.e., the critical orders and LOQ ensures that orders falling in this range are sorted by their timestamps (and a sequencer consults such LOQ instances and sequences them in order of timestamps). So, ME never executes a critical order if it has not executed all the older critical orders which constitutes fairness. When the mid-price moves, LOQ ensures that the messages generated after the movement have lower priority than all the prior messages (due to $I_m$ being the first element of the tuple used by LOQ's priority function), leading to fairness. We conclude this section by reiterating the assumptions that are needed: (i) clock synchronization and (ii) all MPs simultaneously inferring the mid-price movement. Both assumptions are valid in practice with high probability. 


\subsection{Reusing Multicast Tree For Scaling Order Submission Service}

As the number of MPs grows, the exchange's performance degrades as it cannot keep up with offered load. However, we find that even if the cumulative offered load to an exchange stays constant, the growing number of MPs leads to performance degradation because of the increase in \textit{instantaneous} load: The simultaneous order submission from a large number of MPs overwhelms the exchange's ingress. For scaling, the exchange systems usually adopt a sharding approach by partitioning the assets across multiple exchange servers~\cite{cloudex-hotos}. We focus on enhancing the performance of a single exchange server/shard. Our technique is orthogonal to the partitioning based scaling of the exchange. 

With a large number of MPs simultaneously submitting orders to the exchange server, we observe a significant packet loss that leads to decrease in the exchange's throughput and increased latency for orders. We employ a simple strategy: reuse the multicast tree in the reverse direction for proxying the MPs messages to the exchange. The reduced fan-in at the exchange server reduces the packet drops while packet queues form at the tree nodes. We run LOQ at each tree node to service the queues, leading to a significantly high order matching rate of the ME and low latency for orders. 

Furthermore, running LOQ at each proxy node leads to better scheduling compared to running LOQ at just the gateways. A proxy node has messages from several MPs in its queue and can prioritize messages of one MP over the messages of another MP --as long as it does not affect fairness-- where an LOQ at a gateway only has the opportunity of prioritizing messages of an MP over the messages of the same MP. TCP connections are terminated at each tree node to allow the LOQ to reorder the messages. 

\Para{Fairness when using a tree.} Achieving fairness requires that an LOQ instance at each tree node is accompanied by a sequencer instance so that LOQ can operate on the messages of all children nodes fairly. Without a sequencer, a delayed message of an MP may not get assigned its proper priority by LOQ, which would eventually result into safety (fairness) violation at the exchange server. When a sequencer is used at every tree node, it only accounts for messages from its children and not all MPs, so the parameter $n$ changes accordingly in Algorithm~\ref{alg:sequencer}. At each tree node, sequencer discards the incoming dummy messages and each proxy node generates new dummy messages.



\Para{Remarks on other methods for dealing with incast:} Several methods~\cite{homa, protego, breakwater, dctcp} for incast mitigation and overload control have been proposed in the literature. The methods are largely orthogonal to our technique of utilizing the tree in reverse which we can do solely because we have the control of an application that lends us a tree of VMs. Credit-based overload control schemes~\cite{protego, breakwater} are composable to \sysname{}, enhancing \sysname{}'s performance by reducing packet losses/retransmissions. However, a tree further gives us the opportunity to do better scheduling (LOQ) on multiple clients (MPs) data which is not possible without a tree/intermediate nodes and forming the queues only at the order gateways; as would be the case if we use existing techniques. 

%% file: sequencer_protocols.tex
\subsubsection{\textbf{Sequencer Protocol}}

There are $N$ MP VMs where $p_i$ denotes $i^{th}$ MP. Each $p_i$ is connected to the exchange server via a reliable channel (e.g., TCP channel) that provides in-order delivery. Clocks of MPs and the exchange server, where the sequencer runs, are synchronized. MPs generate order messages that are sent to the exchange server. The exchange server maintains a limit order book (LOB) and runs continuous <price, time> priority matching algorithm on all the incoming trade orders. We use $m^{i}_{t}$ to represent an order message from $p_i$ with an order generation timestamp $t$. At the exchange server, messages are fed to the sequencer. A sequencer works in a streaming fashion and takes the messages as input and releases them to the ME. Before releasing a message, it is sequenced as described in the following. 

The sequencer maintains $q : \{m^{i}_{t}\}$ where $q$ is a priority queue with lexicographic ordering on <$t$, $i$> of messages. The sequencer supports \textsc{enqueue} and \textsc{dequeue} operations. 

\textsc{ \textbf{Enqueue}:} As a new message from a $p_i$ arrives at the sequencer, (i) it is added to $q$ and, (ii) if $q$ now contains non-zero number of messages from each $p_i$, dequeue is invoked. 

\textsc{\textbf{Dequeue}:} One message $m^{i}_{t}$ is dequeued from $q$. $m^{i}_{t}$ is considered sequenced at this point and presented to the ME. 

The above sequencer enqueue and dequeue operations provide safety at the cost of degraded liveness. For abetting liveness, each gateway generates a dummy message after every \emph{period} time units on behalf the corresponding $p_i$ if $p_i$ has not generated a message during the last \emph{period} time units. The sequencer dequeue discards a $m^{i}_{t}$ instead of presenting to the ME if $m^{i}_{t}$ is a dummy message where each message contains a field representing whether it is a dummy. 





\begin{algorithm}
\caption{Sequencer Enqueue and Dequeue}
\label{alg:sequencer}
\SetAlgoLined
\KwIn{$n$: Number of total downstreams (MPs) \\
\hspace{3em}$v$: Vector of $n$ lockless FIFO queues \\
\hspace{3em}$q$: Priority queue sorting messages\\
\hspace{4em}lexicographically by $\langle t, i \rangle$ of each message \\
\hspace{3em}$result$: A FIFO queue for sequenced messages
}

\SetKwFunction{Enqueue}{\textbf{Enqueue}}
\SetKwFunction{Dequeue}{\textbf{Dequeue}}

\Enqueue{$m_t^i$}{
    $v[i].enqueue(m_t^i)$\;
}

\Dequeue{}{

    \While{True}{
        $ts \gets \infty $\;
        $ind \gets -1 $\;

        \For{$i \gets 0$ \KwTo $n-1$}{
            \If{$v[i].\text{empty()} = \text{True}$}{
                $ts \gets \infty $\;
                $break$\;
            }

            $m_t^i \gets v[i].top()$\;
            \If{($t = ts$ and $i < ind$) or ($t < ts$)}{
                $ts \gets t $\;
                $ind \gets i$\;
            }
        }

        \If{$ts \neq \infty$}{
            $m_{ts}^{ind} \gets v[ind].dequeue()$\;
            $result.enqueue(m_{ts}^{ind})$\;
        }
    }
}
\end{algorithm}

%% file: loq_protocol.tex
\subsubsection{\textbf{LOQ protocol}}

At a high level, our scheduling scheme installs a special queue (LOQ) at each gateway. LOQ takes the orders as input and schedules them in a way that keeps the matching engine from idling; ME idles if it receives orders that cannot be matched and just needs to be put into the limit order book. Specifically, LOQ categorizes the orders into two classes: In an exchange, orders closer to the mid price -- \emph{critical orders} -- gets matched by the matching engine before the other orders -- \emph{non-critical orders} -- which reside in the LOB and wait for the mid price to move closer. LOQ leverages this domain knowledge to identify and prioritize critical orders over non-critical orders so that the matching engine does not waste time in processing orders that are not going to get matched and only needs to be put in the limit order book while the critical orders remain to be processed.

LOQ scheduling requires two parameters: (i) mid-price $m$ and, (ii) action window $w$. As gateways receive all the market data, they have enough information to infer $m$. Also note that by design of our outbound communication, all the gateways receive each market datum at the same time (with high probability) so all the gateways simultaneously have enough information to infer $m$. Here the action window $w$ indicates \textit{irrationality tolerance}: (i) if an asset is available for purchase for a price $p$ then a buyer can bid on it with a maximum price of $p+w$ and, (ii) if an asset has a highest bid of price $p$ then a seller can ask for a minimum price of $p-w$. Without loss of generality, we can replace $p$ by $m$. The parameter $w$ is configured by the exchange operator. 

An incoming order is categorized as critical if it has a price in the range [$m-w$, $m+w$], otherwise it is non-critical i.e., if a bid has a price in the 
range ($-\infty$, $m - w$) or an ask has a price in the range ($m + w$, $\infty$), then it is non-critical. 

An LOQ is a priority queue that sorts the orders lexicographically by tuple <$\text{I}_{m}$, $c$, $t$> where $\text{I}_{m}$ starts from 0 and is incremented every time $m$ changes, $c$ is 0 if an order is critical otherwise it is 1 and $t$ denotes the generation timestamp of an order. LOQ maintains the global variable $\text{I}_{m}$ while $c$ is calculated per order while enqueuing it. LOQ runs in the gateways where every order generated by the trading algorithm is enqueued to it and a separate thread running the dequeues forwards the orders to the exchange server. The design of the priority function is motivated by the need to not impact inbound fairness as we explain in the following. 

%% file: evaluations.tex
\begin{figure*}[!h]
    \begin{minipage}{0.23\linewidth}
        \centering
        \includegraphics[width=\linewidth]{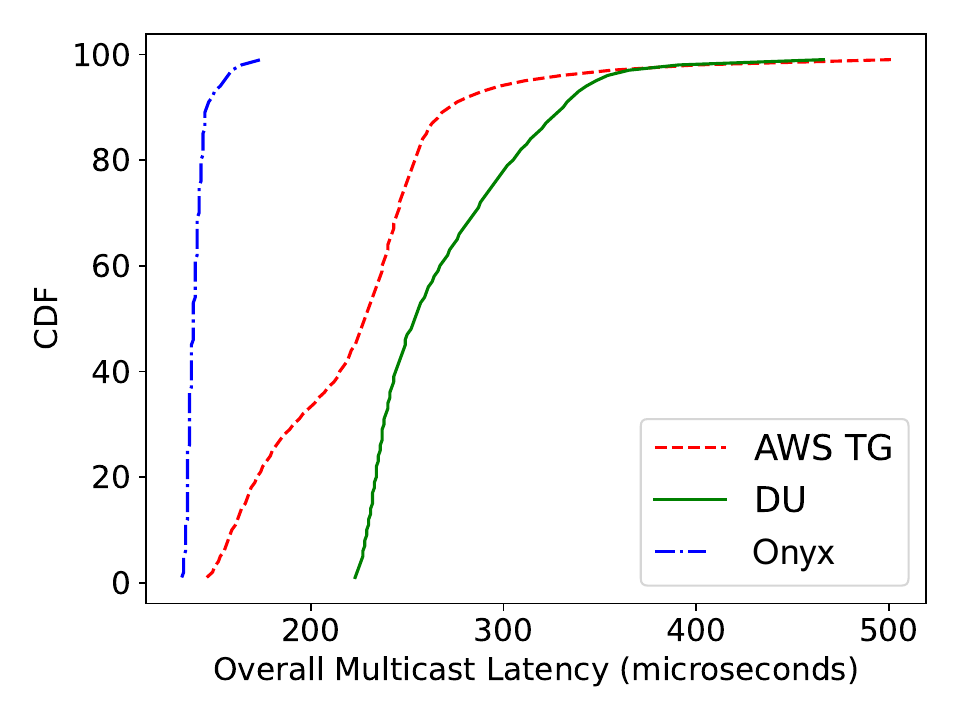}
        \caption{\sysname{} has lower OML than DU, AWS TGW}
        \label{fig:latency:owd_comparison_cdf}
    \end{minipage}
    \hfill
    \nextfloat
    \begin{minipage}{0.765\textwidth}
        \centering
        \begin{subfigure}{0.26\textwidth}
            \centering
            \includegraphics[width=\linewidth]{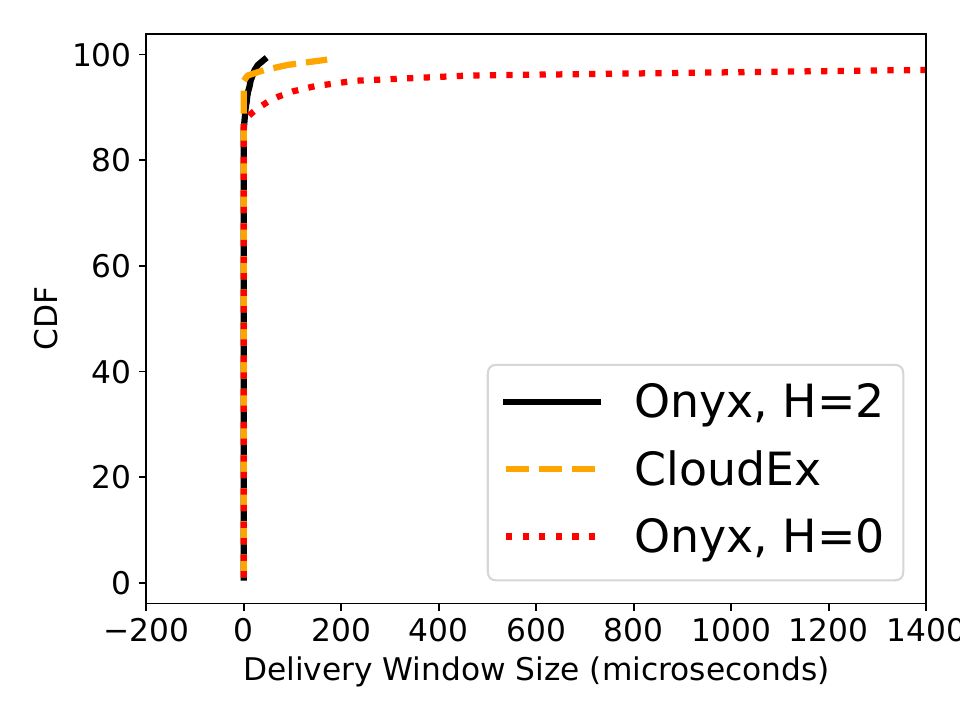}
            \caption{\systemname{} with hedging achieves a narrow DWS}
            \label{fig:sim:holdrelease_h}
        \end{subfigure}\hspace{0.3cm}
        \begin{subfigure}{0.26\textwidth}
            \centering
            \includegraphics[width=\linewidth]{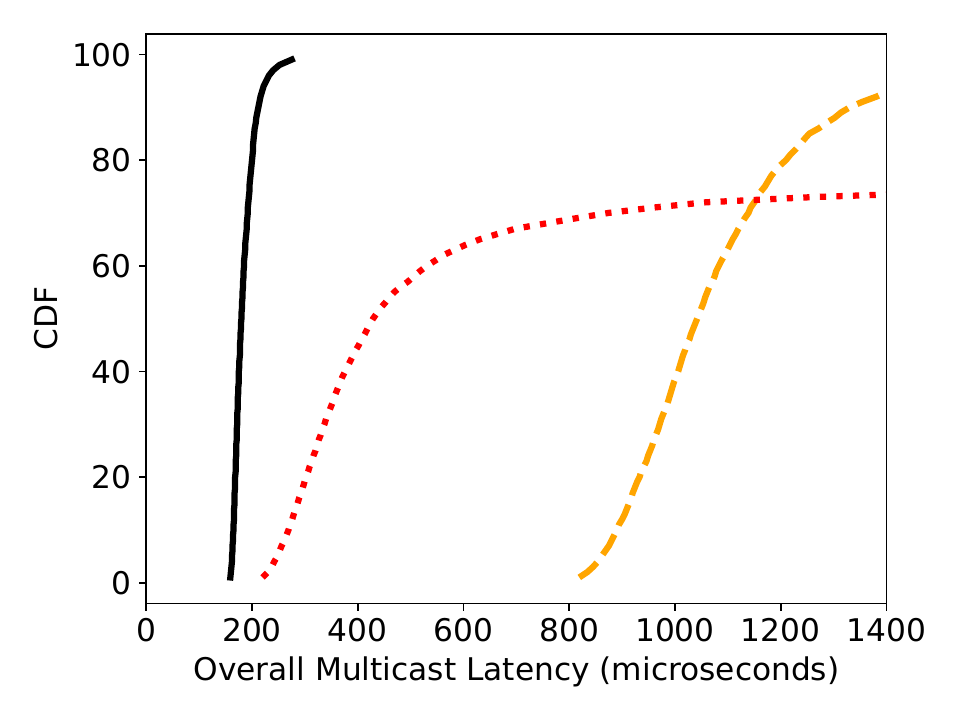}
            \caption{CloudEx shows high OML, like \sysname{} $H=0$}
            \label{fig:sim:owd_holdrelease}
        \end{subfigure}\hspace{0.3cm}
        \begin{subfigure}{0.26\textwidth}
            \centering
            \includegraphics[width=\linewidth]{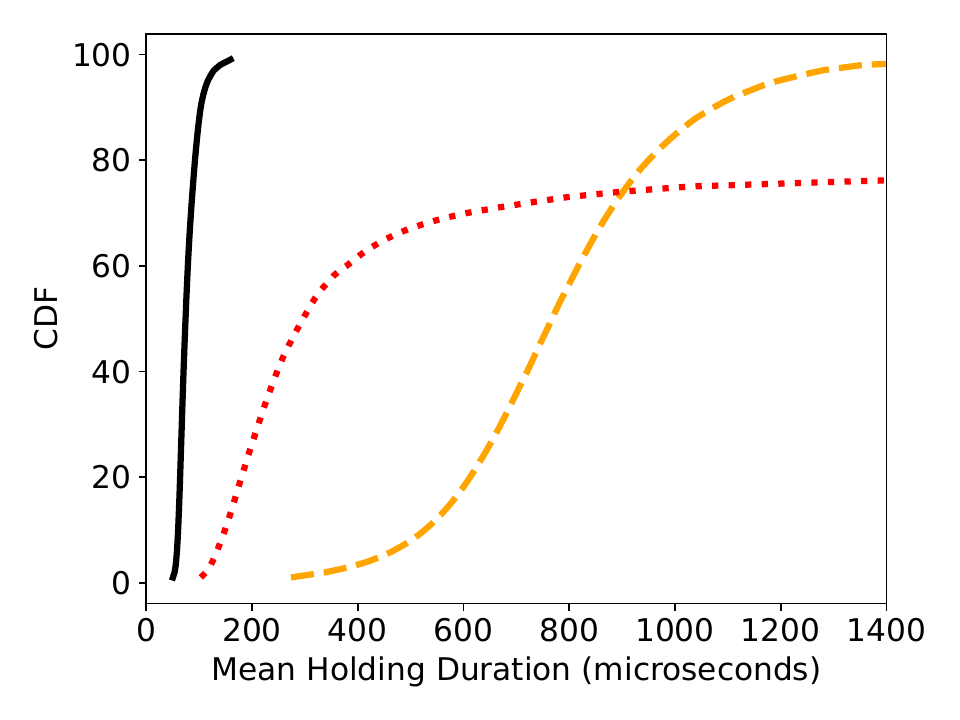}
            \caption{\sysname{} with hedging requires less holding}
            \label{fig:sim:holding_duration}
        \end{subfigure}
        \caption{Outbound fairness and comparison with CloudEx}
    \end{minipage}
\end{figure*}

\section{Evaluation}
\label{evaluations}
We focus on comparison against a prior system, CloudEx~\cite{cloudex-hotos}. As \sysname{} can be viewed as CloudEx augmented with techniques to enhance performance under large number of participants, the comparison with CloudEx shows how much CloudEx can be enhanced. We also present an ablation study as \sysname{} comprises of several techniques put together. System name Jasper and Onyx is used interchangeably in this tech report. 
We answer the following questions.

\begin{enumerate}[labelwidth=!,nosep,label=(\arabic*),leftmargin=*]
    \item How does \sysname multicast latency compare with AWS TGW-based multicast? \S\ref{evals_comp_w_aws_diect}
    \item How much outbound fairness is achieved? \S\ref{evals_sim_delivery}
    \item How does multicast fairness and performance compare to CloudEx? \S\ref{evals_sim_delivery}
    \item How scalable is \sysname? \S\ref{evals_scale}
    \item How does order submission compare to CloudEx? \S\ref{ouch_evals}
    \item How much does LOQ help in enhancing ME's throughput and lowering orders' latency? \S\ref{fancypq_evals}
    \item How much does hedging help? \S\ref{evals_hedging}, \S\ref{sec:evals_rrps}, \S\ref{evals_rcvr_hedging}
    \item How can \sysname{} help DBO? \S\ref{sec:evals_dbo_onyx}
\end{enumerate}

For most of the experiments, we use 100 MPs and a 5K multicast messages per second (MPS) rate. Each MP has a separate VM. For showing scale, some experimemts utilize 1000 MPs and is mentioned in the respective sections. No two MPs share a VM. In practice, multiple MPs could be hosted in a single VM to further enhance scalability. During our evaluation, we define two metrics to quantify the performance of \sysname and the baseline systems. The overall multicast latency (OML) refers to the latency experienced by the last receiver that receives a multicast message. The delivery window size (DWS) is the maximum difference in the latency of any two receivers. We use the Huygens algorithm~\cite{huygens} that synchronizes clocks of all the VMs with a $90p$ offset of $\leq\SI{100}{\nano\second}$. 

\input{evaluations/comparison_summary}
\input{evaluations/sim_delivery_evals}

\input{evaluations/scaling}
\input{evaluations/ouch}
\input{evaluations/itch}
\input{evaluations/onyx_dbo}

%% file: evaluations/comparison_summary.tex

\subsection{Multicast Latency Comparison}
\label{evals_comp_w_aws_diect}

We consider the following baselines and compare their multicast performance with \sysname. 
\begin{itemize}[wide, labelwidth=!,nosep]
    \item Direct Unicast (DU): When conducting multicast, the sender directly sends a copy of a multicast message to each receiver. We use \texttt{io-uring} to create a strong baseline as we observe it reduces the latency by minimizing the overheads of syscalls when doing batch I/0. 
    \item AWS Transit Gateway: AWS-TGW-based multicast is provided by AWS~\cite{aws_tg_multicast}. It requires the sender to send message to a gateway which then replicates and sends one copy to each receiver. It can support at most 100 receivers.
\end{itemize}

Figure \ref{fig:latency:owd_comparison_cdf} shows that \sysname{} distinctly outperforms DU and AWS-TG. This experiment was performed entirely on AWS. The median latency for \sysname{} is \SI{129}{\micro\second} which is $43\%$ lower than the latency of AWS TG (\SI{228}{\micro\second}) and $49\%$ lower than the latency by DU (\SI{254}{\micro\second}). At 90p, \sysname shows $\approx75\%$ lower latency than both DU and AWS TG. Moreover, \sysname{} shows predictable latency as higher percentiles are very close to the median in contrast to the other techniques. 


%% file: evaluations/sim_delivery_evals.tex
\subsection{Outbound Fairness Comparison}
\label{evals_sim_delivery}

We measure the DWS of \sysname and CloudEx to compare their outbound fairness performance. A smaller DWS indicates a higher level of outbound fairness (i.e., simultaneous delivery).


Figure \ref{fig:sim:holdrelease_h}  shows that \systemname{} can achieve a DWS of $\leq\SI{1}{\micro\second}$ at very high percentiles (up to $92p$). Without proxy hedging, the DWS becomes larger at earlier percentile ($87^{th}$). CloudEx achieves fairness but at the cost of high OML. 

Figure \ref{fig:sim:owd_holdrelease} shows that OML increases significantly for CloudEx and \systemname{} with no hedging. In these systems, the deadlines calculated by the hold-and-release are far into the future to cover the high latency variance. This leads to high holding duration at the receivers for CloudEx and \sysname{} without hedging as shown in figure \ref{fig:sim:holding_duration}, increasing OML. 

We do not include a comparison with DBO here as DBO changes the definitions of fairness to not rely on simultaneous delivery. However, the latency of DBO's market data multicast would be slightly worse than a Direct Unicast (DU) approach discussed in~\S\ref{evals_comp_w_aws_diect} as it does batching and pacing of multicast packets. \sysname{} outperforms DU.

%% file: evaluations/scaling.tex
\subsection{Scaling \sysname{} Multicast}
\label{evals_scale}
\begin{figure}[!h]
    
  \begin{minipage}{0.45\linewidth}
      \centering
      \vspace{-0.3cm}
      \includegraphics[width=1\linewidth]{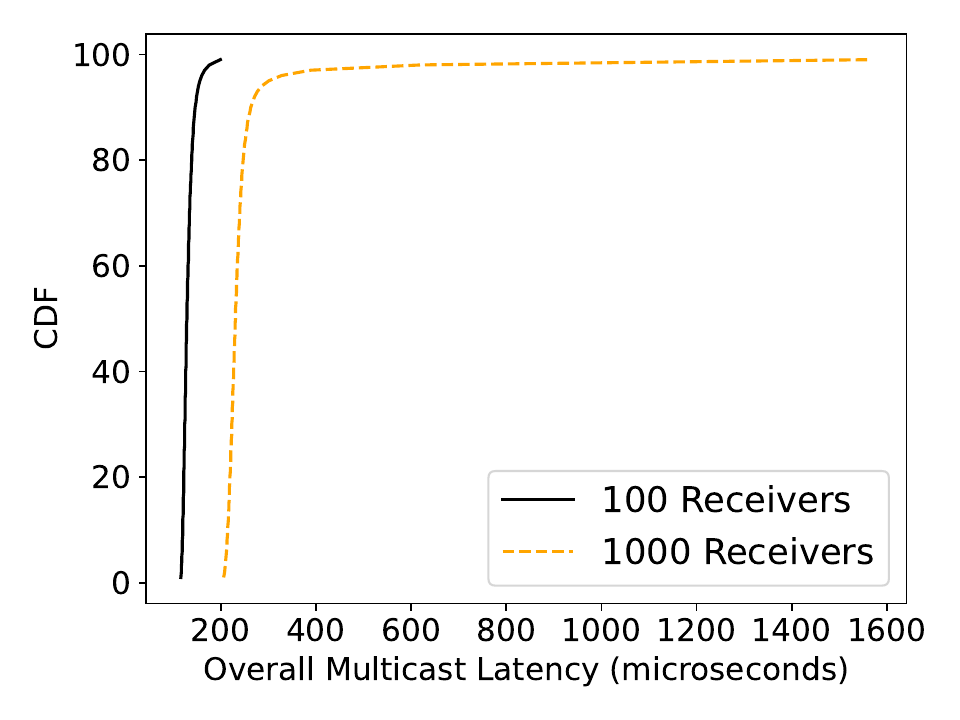}
      \vspace{-0.7cm}
     \caption{\sysname{} scales well with OML $\leq $250$\SI{}{\micro\second}$}
     \label{fig:scale:receivers}
     \vspace{-0.5cm}
  \end{minipage}
  \hfill
  \begin{minipage}{0.45\linewidth}
    \centering
    \vspace{-0.3cm}
    \includegraphics[width=1\linewidth]{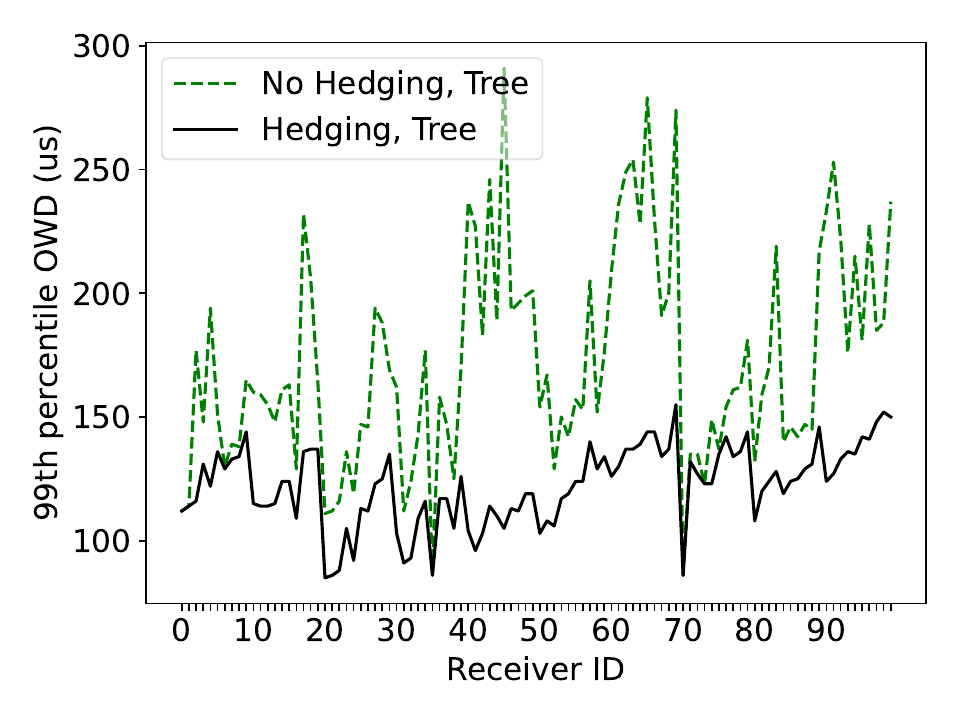}
    \vspace{-0.7cm}
    \caption{OWD per receiver improves with hedging}
    \label{fig:ablation:spatial_owd_tree_99p}
    \vspace{-0.5cm}
  \end{minipage}
\end{figure}

\Para{Scaling N to 1000.} In contrast to the previous fair multicast solutions that only work with a few 10s of receivers (e.g.,~\cite{dbo_prateesh,octopus_google,aws_tg_multicast}), \sysname aims to implement a more scalable multicast service which can support 1000 receivers and potentially even more. Figure \ref{fig:scale:receivers} plots \sysname's multicast latency for 100 receivers and 1000 receivers. Our overlay techniques for reducing latency enable \sysname to keep a graceful latency growth as the number of receivers increase.


\Para{{Fairness for an increasing number of receivers.}} When employing hold-and-release, a median DWS of $\leq\SI{1}{\micro\second}$ is achieved for $N=1000$. However, the highest percentile at which DWS stays $\leq\SI{1}{\micro\second}$ decreases as $N$ increases. We define the probability of fairness ($P$($F$)) as the highest percentile where DWS is $\leq\SI{1}{\micro\second}$. Table \ref{tab:fairness_receivers} shows OML and $P$($F$) for $N$ = 100 and 1000. 



\Para{Throughput.} \sysname can utilize $\approx$80\% of the egress bandwidth of a \texttt{c2d-highcpu-8} VM on GCP while maintaining negligible packet losses. The specified VM has an egress bandwidth of 16Gbps. With 466B packets, we achieve a multicast message rate of 350K MPS, without proxy hedging. The $50p$ OML stays the same as we increase the message rate from 5K to 350K, but $99p$ increases $13\%$. In the experiment, we have 100 receivers and 10 proxies, with no proxy hedging. The sender and each proxy replicate each message $F=10$ times hence, with 350K messages per second, the sender generates 3.5M packets per second, utilizing 81.5\% of the available egress. The proxy hedging would reduce the multicast rate according to the factor $H$ as each proxy replicates each multicast message $(H+1)*F$ times, e.g., with $H=1$, we achieve 175K messages per second. 

%% file: evaluations/ouch.tex
\subsection{Orders Submission Performance}
\label{ouch_evals}

\begin{figure}[bh]
\vspace{-0.2cm}
\centering
\includegraphics[width=0.35\textwidth]{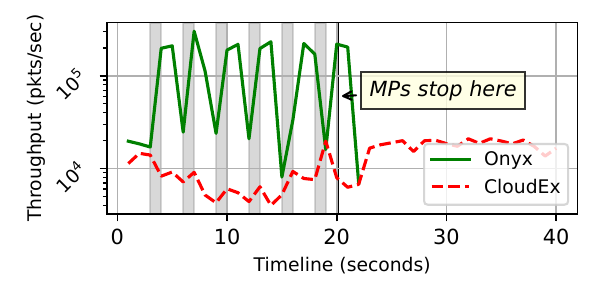}
\vspace{-0.5cm}
\caption{Onyx achieves high order matching rate in comparison to CloudEx.}
\vspace{-0.3cm}
\label{fig:comparison:onyx_dbo_cloudex_thpt}
\end{figure}

Figure \ref{fig:comparison:onyx_dbo_cloudex_thpt} shows the order matching rate of the exchange server for three systems. Onyx outperforms CloudEx by atleast an order of magnitude. 100 MPs cumulatively generate 100K orders per second each where periodic bursts occur (shown as shaded regions). The order generation rate becomes $20\times$ during the bursts. MPs stop generating the orders after 20s. Onyx is able to keep up with the offered load and finishes processing the orders right after the MPs stop. Onyx achieves higher order matching rate when bursts occurs due to LOQ. CloudEx achieves significantly low order matching rate and builds up queues that are processed even after the MPs stop. The median latency of orders with Onyx is $\approx97\%$ lower compared to CloudEx.

\begin{figure}[!t]
    \vspace{-0.3cm}
    \begin{minipage}{0.23\textwidth}
        \centering
        \includegraphics[width=1\linewidth]{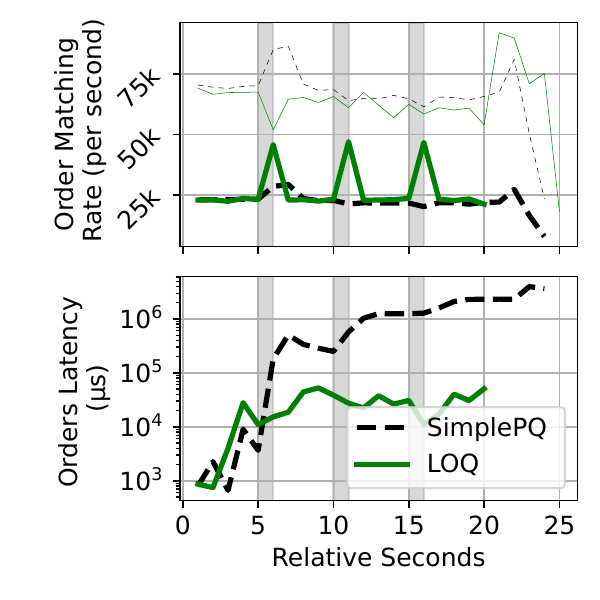}
        \vspace{-0.7cm}
        \caption{LOQ effectively handles bursts (shaded). Thin lines = Input to ME}
        \vspace{-0.5cm}
        \label{fig:fancy_v_simple_thpt_lat}
    \end{minipage}
    \hfill
    \begin{minipage}{0.23\textwidth}
        \centering
        \includegraphics[width=1\linewidth]{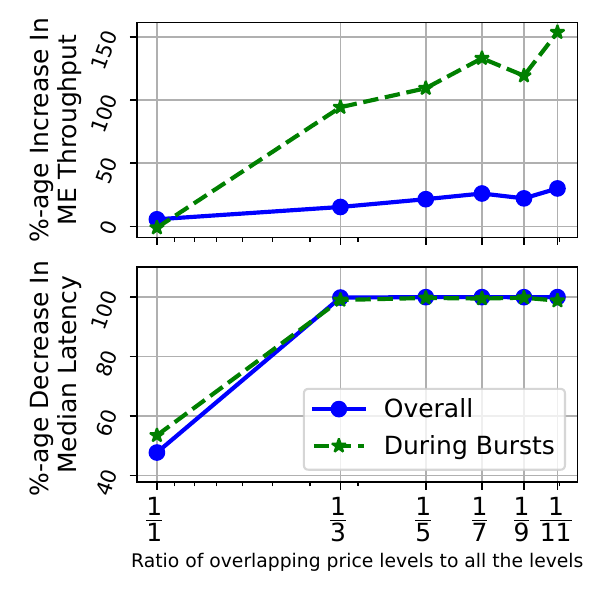}
        \vspace{-0.7cm}
        \caption{Benefits of LOQ increase as the book depth increases.}
        \vspace{-0.5cm}
        \label{fig:fancy_v_simple_book_depth}
    \end{minipage}
\end{figure}

\Para{Limit Order Queue (LOQ) Performance}
\label{fancypq_evals}
We compare LOQ to a first-in-first-out queue, which we refer to as SimplePQ. Unlike LOQ, SimplePQ does not differentiate between critical and non-critical orders. In this experiment, we have 10 market participants (MPs), each submitting 7K orders per second. Two intermediate proxies, running LOQ or SimplePQ, relay orders from the MPs to the ME. Order bursts occur, with each MP doubling their order submission rate every \SI{5}{\second}. Bursts lead to queue build ups at the proxies, where LOQ/SimplePQ operate. As shown in Figure \ref{fig:fancy_v_simple_thpt_lat}, LOQ outperforms SimplePQ in ME's order matching rate and matched orders' latency. Thin lines represent the orders received per second by ME which are similar (but noisy) for both SimplePQ and LOQ. Thick lines represent order matching rate and orders latency in their respective plots showing LOQ's better performance. 

The advantages of LOQ stem from the presence of non-critical orders, which are kept in the gateways and proxies' queues longer, allowing the available egress to be used for critical orders. We examine the benefits of LOQ under different ratios of critical to non-critical orders. In this experiment, MPs uniformly sample bid/ask prices from a predefined range to generate orders. We vary the portion of the bid price range that overlaps with the ask price range. These overlapping segments result in orders that get matched at ME, while other orders remain in the LOB. Figure \ref{fig:fancy_v_simple_book_depth} shows that, as the ratio of overlapping price levels to the total number of levels decreases, the benefit of LOQ increases. We plot the metrics for the bursts duration which are calculated from the orders matched during the bursts that we inject. A ratio of \(x/y\) indicates that there are \(y\) total price levels from which MPs sample prices to generate orders, but only \(x\) price levels lead to immediate matches at ME. As \(y\) increases while \(x\) remains constant, the number of critical orders decreases. LOQ identifies critical orders and prioritizes them for processing by the ME over non-criticals. Consequently, LOQ demonstrates a 150\% increase in ME throughput and a 99.9\% reduction in latency for matched orders when \(y=11\).\footnote{\(y\approx10\) is commonly used in the literature~\cite{liquidity_supply_electronic_markets}} We repeat one experiment and increase MPs from 10 to 1000 with $x=1, y=7$ and observe that LOQ distinctly outperforms SimplePQ: 98\% decrease in the latency and 10\% (86\%) increase in overall (during bursts) matching rate.

\begin{figure}[!b]
\vspace{-0.2cm}
\centering
\includegraphics[width=0.35\textwidth]{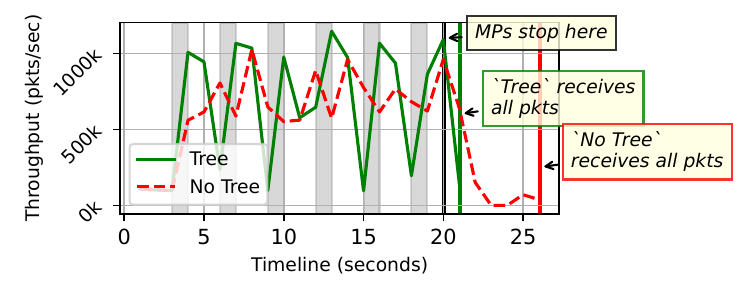}
\vspace{-0.5cm}
\caption{A tree enhances an exchange's throughput.}
\vspace{-0.3cm}
\label{fig:comparison:tree_notree_thpt}
\end{figure}

\Para{Using Proxy Tree for Orders Submission:} To conduct an ablation study of the proxy tree's benefit in the order submission, we run an experiment without LOQ and sequencer, and measure the number of packets received per second by the exchange server with and without a tree. Figure~\ref{fig:comparison:tree_notree_thpt} plots the throughput achieved with 100 MPs. 100 MPs send at the cumulative base rate of 100K packets/second, but create periodic bursts that reach $20\times$ higher than the base rate. By using the proxy tree, \sysname can improves the overall throughput by $\sim22\%$ on average and by $\sim75\%$ during the bursty periods.



\begin{figure}[!b]
\vspace{-0.2cm}
\centering
\includegraphics[width=0.35\textwidth]{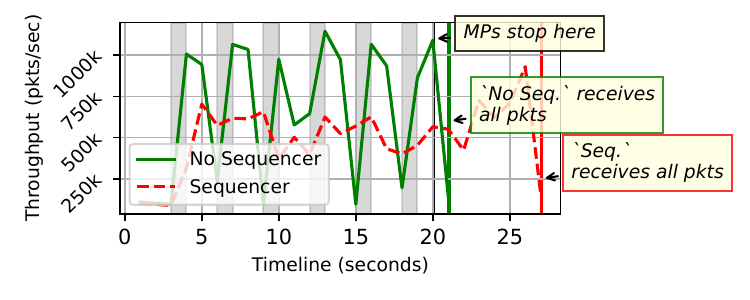}
\vspace{-0.5cm}
\caption{Sequencer has a significant overhead. }
\vspace{-0.3cm}
\label{fig:comparison:seq_noseq_thpt}
\end{figure}

\Para{Sequencing Overhead:} We perform an experiment with 100 MPs (without LOQ) to study the throughput with and without our sequencer. We use the same load generator as above and notice an overall 25\% decrease in the throughput of the exchange when using the sequencer. The throughput decrease is traded for improving inbound fairness. On the other hand, our LOQ protocol improves the \sysname throughput by assigning priorities to the packets and facilitate the order matching at the exchange. As a result, the incorporation of the proxy tree and LOQ protocol compensates for the sequencer overheads and enables \sysname to outperform CloudEx (Figure~\ref{fig:comparison:onyx_dbo_cloudex_thpt}).


\Para{Remarks on inbound fairness: } By design, our sequencer (w/o LOQ) ensures 100\% inbound fairness. One of the assumptions for LOQ to respect fairness is that all MPs/gateways infer the mid-price movement at the same time which is provided by outbound fairness with a high probability. Therefore, the probability of achieving inbound fairness while employing LOQ can only be as high as that of outbound fairness which, we show in \S\ref{evals_sim_delivery} and \S\ref{evals_scale}, is achieved by 90\% probability. In \S\ref{evals_rcvr_hedging}, we will continue to show the probability of outbound fairness can be optimized to 99.9\%, which is much better than CloudEx. 


%% file: evaluations/itch.tex
\begin{figure*}[h]
    \centering
    \begin{minipage}{0.2\textwidth}
        \centering
        \begin{tabular}{ccc}
        \toprule
        N & P(F) (\%) & OML \\
        \midrule
        \rowcolor{lightgray}
        100  & 92 & 175 \\ 
        1000 & 89 & 238 \\ 
        \bottomrule
        \end{tabular}
        \captionof{table}{\sysname{} achieves fairness for a large $N$ with low OML($\SI{}{\micro\second}$).}
        \label{tab:fairness_receivers}
    \end{minipage}%
    \hfill
    \nextfloat
    \begin{minipage}{0.8\linewidth}
        \begin{subfigure}[]{0.3\textwidth}
        \includegraphics[width=\linewidth]{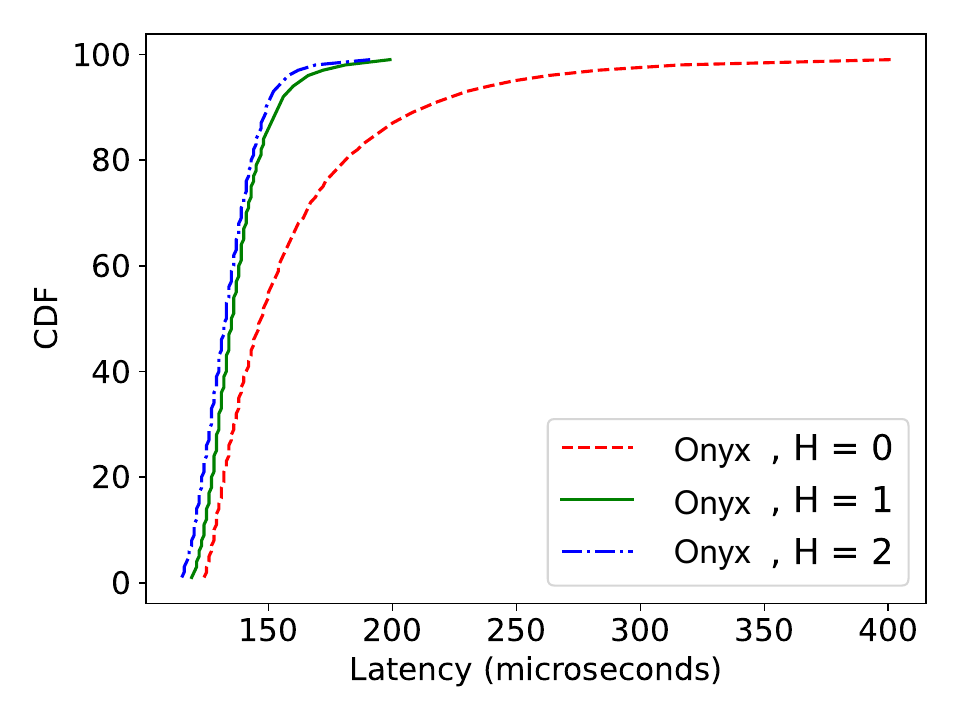}
        \caption{Hedging reduces OML}
        \label{fig:hedging:increasing_h_owd}
        \end{subfigure}\hspace{0.3cm}
        \begin{subfigure}[]{0.32\textwidth}
        \includegraphics[width=\linewidth]{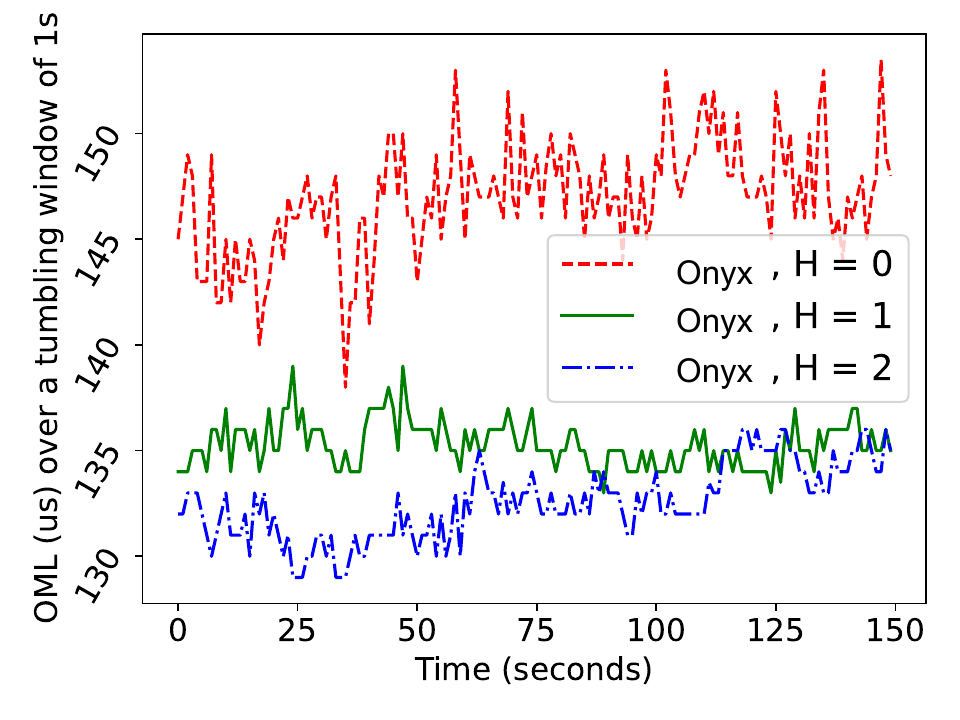}
        \caption{Low temporal variance}
        \label{fig:hedging:temporal_consistency}
        \end{subfigure}\hspace{0.3cm}
        \begin{subfigure}[]{0.28\textwidth}
        \includegraphics[width=\linewidth]{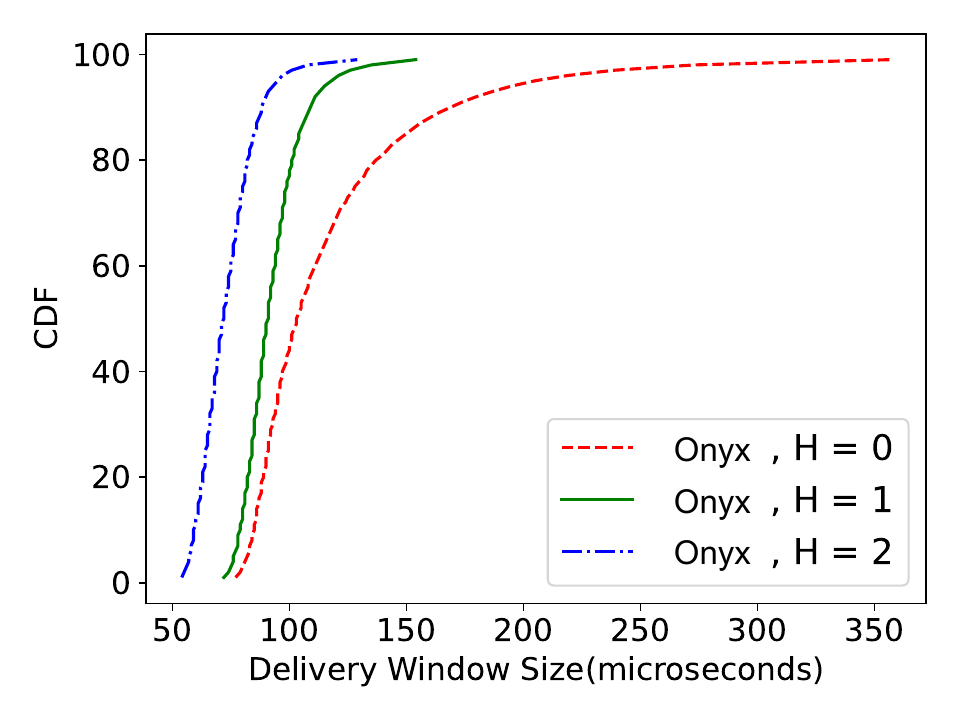}
        \caption{Hedging reduces DWS}
        \label{fig:hedging:spatial_consistency}
        \end{subfigure}
        \vspace{-0.3cm}
        \caption{Evaluating Proxy Hedging}
    \end{minipage}
\end{figure*}

\subsection{Outbound Communication Techniques}

\input{evaluations/hedging_evals}
\input{evaluations/dyn_relations}
\input{evaluations/receiver_hedging}

\begin{table}[!h]
\centering
\scriptsize
\vspace{-0.24cm}
\begin{minipage}[t]{0.24\textwidth}
\centering
\begin{tabular}{c|c|c|c|c}
\toprule
\textbf{Setup} & \multicolumn{2}{c|}{\textbf{On AWS}} & \multicolumn{2}{c}{\textbf{On GCP}} \\ 
 & \textbf{50p} & \textbf{90p} & \textbf{50p} & \textbf{90p} \\
\midrule
\rowcolor{lightgray}
\makecell{\textbf{Improve.}\\\textbf{(\%)}} & 15.12 & 69.35 & 9.83 & 9.18 \\
\bottomrule
\end{tabular}
\caption{\%-age improvement in OML due to RRPS}
\label{tab:dyn_vs_fix_oml}
\end{minipage}%
\hfill
\hfill
\begin{minipage}[t]{0.23\textwidth}
\centering
\begin{tabular}{cccc}
\toprule
$N$ & RH & \multicolumn{2}{c}{OML (\SI{}{\micro\second})} \\
& & 50$p$ & 99$p$ \\
\midrule
\rowcolor{lightgray}
100 & No & 139 & 248 \\ 
200 & No & 141 & 243 \\ 
\rowcolor{lightgray}
100 & Yes & 99 & 146 \\
\bottomrule
\end{tabular}
\caption{Latency impact due to Receiver Hedging (GCP)}
\label{tab:rcvr_hedging_lat}
\end{minipage}
\vspace{-1cm}
\end{table}

%% file: evaluations/hedging_evals.tex
\subsubsection{\textbf{Proxy Hedging}}
\label{evals_hedging}



\Para{\\Reduced overall multicast latency.} Figure~\ref{fig:hedging:increasing_h_owd} compares the OML CDFs under different hedging factors ($H=0,1,2$). $H=0$ represents the case when proxy hedging is not enabled. Compared with $H=0$, $H=1$ yields distinct latency reduction as the CDF curve is shifted to the left. The latency reduction is marginal as H increases from 1 to 2. We show later that setting $H=2$ yields better DWS than $H=0,1$.

\Para{Reduced temporal and spatial latency variance.} We calculate the median OML over a tumbling window of 5000 messages to study the temporal latency variance during the multicast. Figure~\ref{fig:hedging:temporal_consistency} shows \systemname{} with hedging exhibits lower temporal latency variance compared with the non-hedging scheme ($H=0$). Enhancing the temporal predictability of latency, even for short time-steps, is beneficial in multiple ways, particularly in the hold-and-release mechanism, which relies on latency estimates to determine message deadlines. 

Figure~\ref{fig:hedging:spatial_consistency} compares the CDFs of DWS for \systemname{} with different hedging factors. Hedging substantially reduces the spatial latency variance (i.e., it shrinks the DWS of multicast messages). The 99th percentile delivery window size is $\sim$\SI{350}{\micro\second} with $H=0$ but $\sim$\SI{150}{\micro\second} with $H=1$. However, as $H$ grows from 1 to 2, the window size reduction, although it still exists, becomes less distinct. 

\Para{Reduced OWDs for each receiver.} Proxy hedging improves OML as the OWD to each receiver is reduced. Figure \ref{fig:ablation:spatial_owd_tree_99p} shows OWD for each receiver with ($H=2$) and without hedging. 

%% file: evaluations/dyn_relations.tex
\subsubsection{\textbf{Round Robin Packet Spraying}}
\label{sec:evals_rrps}

We evaluate the impact of RRPS on the multicast latency. We utilize a base multicast message rate of 10K MPS, 125 receivers and a burst of \SI{1}{\second} occur every \SI{2}{\second}. During bursts, the message rate becomes $15\times$. Table \ref{tab:dyn_vs_fix_oml} shows ulticast latency decreases when RRPS is used. On GCP, we see an OML reduction of $\approx10\%$, however, on AWS the reduction can approach $\approx70\%$. With low message rates, the reduction is not significant: $\leq5\%$ with message rates $\leq100K$ MPS. This happens because the inter-packet duration is small enough that any transient latency spike is not able to impact consecutive messages so RRPS does not help much by re-routing messages. 


RRPS show improvements in OML when message bursts are introduced as it can distribute the bursts among several network paths. Without bursts, the OML reduction still happens but is less than $10\%$ on both AWS and GCP. 

%% file: evaluations/receiver_hedging.tex

\subsubsection{\textbf{Receiver Hedging}}
\label{evals_rcvr_hedging}
The receiver hedging improves OML (Table~\ref{tab:rcvr_hedging_lat}) and achieves fairness (i.e., DWS $\leq\SI{1}{\micro\second}$ at $99.9p$, when hold-and-release is used). Receiver hedging is not utilized in any experiments other than Table~\ref{tab:rcvr_hedging_lat}. This technique improves \sysname{} performance at the cost of one extra VM per MP. Doubling the amount of receiver VMs ($N$) sometimes also require restructuring the proxy tree which incurs extra dollar cost: 10 proxies are required for $N$=100, 14 proxies for $N$=200.


%% file: evaluations/onyx_dbo.tex
\subsection{Onyx and DBO}
\label{sec:evals_dbo_onyx}

\input{dbo}

%% file: dbo.tex
DBO is a recent cloud exchange that achieves fairness regardless of latency fluctuations and without clock synchronization. It does so by proposing a new fairness metric that prioritizes orders based on the response time of traders (i.e., the time to make a trade in response to a piece of market data), rather than the time at which orders were submitted. Onyx can be viewed as a network layer that is complementary to DBO's new semantics at the application layer. In particular, Onyx can help DBO scale out to a large number of participants, particularly in the following three ways. One, Onyx's low-latency and scalable multicast tree (and \emph{hedging}) can help with disseminating market data in DBO.  Two, Onyx's use of a tree in reverse for inbound order submission can also help DBO deal with incast-type conditions during order submissions from a large number of participants. Three, because DBO does not enforce simultaneous delivery of data, DBO has an extra constraint that no MPs should directly or indirectly talk to each other as it can violate DBO's fairness guarantees. By leveraging Onyx's simultaneous multicast in DBO, this constraint can be lifted.

%% file: related_work.tex
\section{Related Work}
\label{related_work}
We have already discussed the recent exchanges (CloudEx~\cite{cloudex-hotos}, DBO~\cite{dbo_prateesh}) and overload control schemes~\cite{protego, breakwater, homa}. 


\Para{Multicast:} Prior works on application level multicast~\cite{nice_multicast, delay_and_delay_variation_multicast, iterative_algorithm_multicast, narada, delay_multicast_mesh_overlays} mainly focus on finding optimal paths in a network (or overlay mesh) using cost models for network \emph{links} that capture heterogeneous link bandwidth or latency characteristics. \systemname{} focuses on (i) minimizing the delays incurred at \emph{hosts} while transmitting multiple message copies, and (ii) achieving a small latency difference across receivers. On the other hand, switch-based multicast~\cite{switch_multicast1, switch_multicast2} is not available to cloud tenants due to scalability issues~\cite{elmo}. 

\Para{Collective Communication:} Collective communication~\cite{sigcomm21-hoplite, collective_communication_1, collective_communication_2} uses overlay trees for broadcast and all-reduce~\cite{all_reduce}, but typically supports fewer than 100 receivers. For example, Hoplite~\cite{sigcomm21-hoplite} optimizes bandwidth for 10s of nodes using a tree. \systemname{} focuses on minimizing latency and variance, handling bursts, and achieving fairness at scale while not having an \emph{aggregation} mechanism for trade orders. 

%% file: conclusion.tex
\section{Conclusion}


This paper introduces \systemname{}, which provides networking support for scalable financial exchanges. \systemname{} systematically optimizes both inbound (order submission) and outbound (market data delivery) workflows of the exchange system. Our evaluation shows that \systemname{} outperforms a prior system CloudEx, in terms of fairness, throughput, and latency as well as outperforms AWS TGW-based multicast. This work does not raise any ethical issues. 




%% file: appendix.tex
\appendix

\begin{table*}

\subcaptionbox{N=10}{
\begin{tabular}{ccc}
\toprule
D & F & OML \\
\midrule
1 & 10 & \cellcolor{green}66 \\
2 & 4 & 88 \\
\bottomrule
\end{tabular}
}
\hfill
\subcaptionbox{N=100}{
\begin{tabular}{ccc}
\toprule

D & F & OML \\
\midrule
1 & 100 & 351 \\

2 & 10 & \cellcolor{green}139 \\

3 & 5 & 141 \\
\bottomrule
\end{tabular}
}
\hfill
\subcaptionbox{N=1000}{
\begin{tabular}{ccc}
\toprule
D & F & OML \\
\midrule
2 & 32 & 282 \\

3 & 10 & \cellcolor{green}217 \\

4 & 6 & 226 \\
\midrule
\end{tabular}
}

\caption{\textmd{Median values of overall multicast latency (OML), in \SI{}{\micro\second}, for different depth ($D$), fanout ($F$), and receivers ($N$).}}
\label{tb:micro-bench}
\end{table*}

\section{Deciding D and F for multicast tree}
\label{app:decide_d_and_f}

To understand how the end-to-end latency varies as $D$ and $F$ grow, we conduct a series of experiments with various numbers of receivers ($N=10,100,1000$). Table~\ref{tb:micro-bench} displays the latencies for different configurations of <$D, F$> for a given number ($N$) of receivers. We see that the latency reaches a minimum point at different depths $D$ for different values of $N$. At a small scale ($N=10$), increasing $D$ does not bring latency benefits. As the scale grows from $N=10$ to $N=100,1000$, the benefits of increasing $D$ while reducing $F$ grow because the reduced message replication delay and transmission delay outweigh the added overhead of new hops in the path of messages. However, as $D$ goes beyond a certain threshold (e.g., $D$ grows larger than 3 when $N=1000$), the latency no longer decreases but gets worse. Based on our experiments, we find fixing $F=10$ usually leads to a desirable $D$ to generate a multicast tree with low latency. We establish our heuristic rule to construct the multicast tree as follows: Given $N$ receivers, we fix $F=10$ and then derive $D=[log_{10}N]$ (round to the nearest integer). We also tested a more sophisticated alternative mechanism described in \cite{nsdi16_ernest}, and observe no significant performance boost compared to our heuristic. 

In our testing, we find that the variance in cloud-host VM performance~\cite{lemondrop} imposes the great challenge to differentiate the ``optimal'' values of $D$ and $F$ from values picked using the aforementioned heuristic. We observe through repeated experimentation that minor changes in $D$ and $F$ do not show a significant performance difference with high statistical confidence. So it is sufficient to select $D$ and $F$ in the neighborhood of the optimal value, which is why our heuristic is effective. We further find that a unit increment in $D$ comes at the added latency of $30\pm 10\mu\/s$ and a unit increment in $F$ adds $2.7 \pm 0.9 \mu\/s$ per layer.\footnote{Using our high-performance implementation on a c5.2xlarge VM in AWS. Message size is 460B. } A tree constructed using linear models based on these unit increments performs comparably to the tree constructed using our heuristic of maintaining $F$ close to 10 and $D=[log_{10}N]$. An example scenario for deciding tree structure: For $N=100$, $F=10$, $D$ would be 2. For $N=200$, $F=10$, $D$ would still be 2 and F would need to adjusted accordingly to support all 200 receivers. So F would come around to be be 14 as $14^{2} \approx 200$.

\section{Scalable Simultaneous Delivery}
\label{app:sim_delivery}

Financial trading needs to ensure the fairness of the competition. Fairness in data delivery~\cite{cloudex-hotos,octopus_google} means that the market data from the exchange server should be delivered to every MP at the same time so that an MP may not gain an advantage over the others during the competition. While there are also some recent works trying to alter the definition of fairness~\cite{dbo_prateesh,libra}, these variant definitions require more research before they are confidently adopted by the exchanges. Therefore, \systemname{} uses the original definition of fairness employed by on-premises financial exchanges. In short, in \systemname{} the fair delivery of data means simultaneous delivery of market data to all the multicast receivers. 

\emph{Realizing perfect simultaneous data delivery to multiple receivers is theoretically unattainable}~\cite{two_generals_problem}. Nevertheless, \systemname{} tries to empirically minimize the spatial (i.e., across receivers) variance of the latency of messages. Our hedging design (\S\ref{sec:vm_hedging}) has created favorable conditions to minimize the spatial variance. By using hedging, \systemname{} can achieve consistently low variance for each receiver over time, and the spatial latency variance across receivers is kept low. Beyond this, we employ a \emph{hold-and-release} mechanism (by using synchronized clocks) to eliminate the residual spatial variance at the end hosts and enforce simultaneous delivery across receivers. The \emph{hold-and-release} mechanism was introduced by CloudEx~\cite{cloudex-hotos} but it does not scale well and leads to high end-to-end latency (\S\ref{evals_sim_delivery}). We describe a modified \emph{hold-and-release} mechanism that helps us scale further, while maintaining a low end-to-end multicast latency. 

\Para{Hold \& Release mechanism.}
To implement the \emph{hold-and-release} mechanism, \systemname{} leverages the accurate clock synchronization algorithm, Huygens~\cite{huygens}, to synchronize the clocks among the sender and receivers. Receivers keep track of the one-way delay (OWD) of the messages received from the sender. Each receiver takes the 95th percentile of its OWD records at regular intervals and sends the results back to the sender using an all-reduce mechanism explained later. The sender calculates the maximum of these OWDs called Global OWD for messages using the gathered statistics: $\mathbf{OWD_G = \max_{i}(OWD_i)} $ where $\mathbf{OWD_i}$ is the OWD estimate reported by the $i$-th multicast receiver.

Once an $\mathbf{OWD_G}$ has been calculated by the multicast sender, it attaches deadlines to all outgoing messages. A deadline is calculated by adding $\mathbf{OWD_G}$ to the current timestamp when sending a message. Upon receiving a message, a receiver does not process it until the current time is equal to (or exceeds) the message's deadline. This mechanism leads to almost simultaneous delivery, modulo clock sync error. In \S\ref{sec:trust_model}, we discussed possible security mechanisms to ensure that a receiver (an MP) waits until its deadline to process a message, when it's in the MP's self interest not to wait. 

\Para{Deadlines all-reduce.} In the design of \systemname{}'s \emph{hold-and-release} mechanism, all the receivers have to send estimates of the OWD they experience back to the sender so that the sender can estimate the deadlines for the subsequent messages to multicast. If all the receivers send their estimated OWD directly to the sender, incast congestion occurs at the sender, leading to increased message drops.\footnote{We measure that more than 20\% of the packets are dropped when 100 receivers attempt to send OWDs back to the sender simultaneously. }  To avoid the high volume of incast traffic, we reuse the multicast tree to aggregate the OWD estimates in an all-reduce manner~\cite{all_reduce}. Specifically, each receiver periodically sends its OWD estimate to its parent proxy. As each proxy has a limited number of children, we do not risk in-cast congestion here. Each proxy (i) continuously receives estimates from its children; (ii) periodically takes the maximum of all the received estimates, ignoring some children OWDs if they have not yet sent in an estimate, and; (iii) sends the max value to its parent proxy. In this way, the sender at the root calculates deadlines for messages by only receiving the aggregated estimates from the first proxy layer instead of all the receivers.

\section{Hedging Analysis Via Monte Carlo}
\label{app:hedging_analysis}
\begin{figure*}[h]
  \begin{subfigure}[]{0.3\textwidth}
        \includegraphics[width=\textwidth]{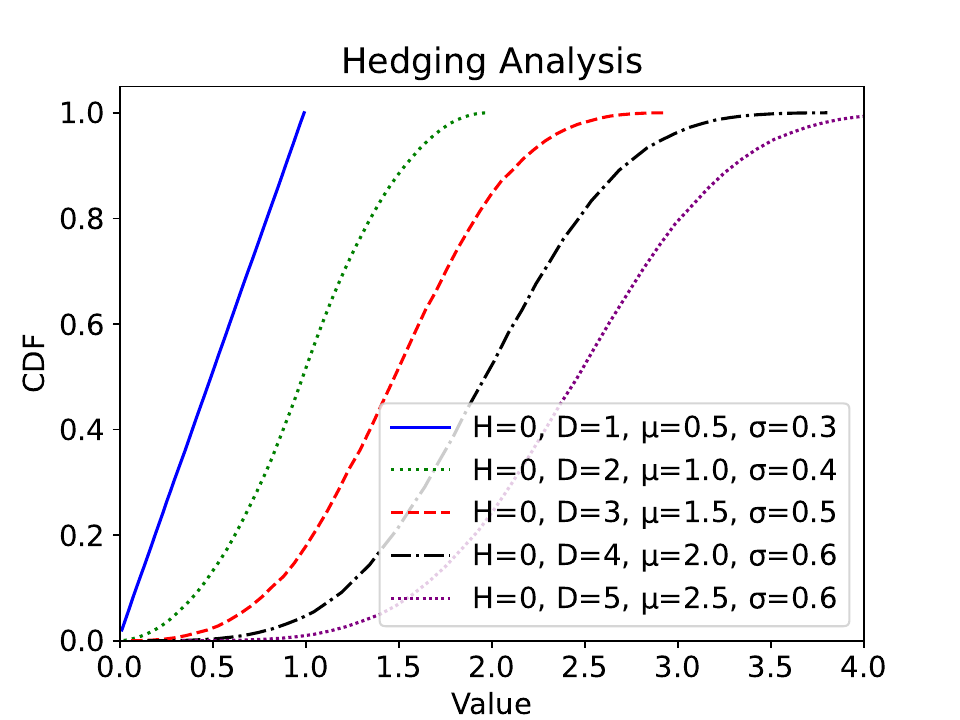}
        \caption{With no hedging, the depth of a tree makes the latency and its variance worse.}
        \label{fig:rv:h1_d}
  \end{subfigure}\hspace{0.3cm}
  \begin{subfigure}[]{0.3\textwidth}
        \includegraphics[width=\textwidth]{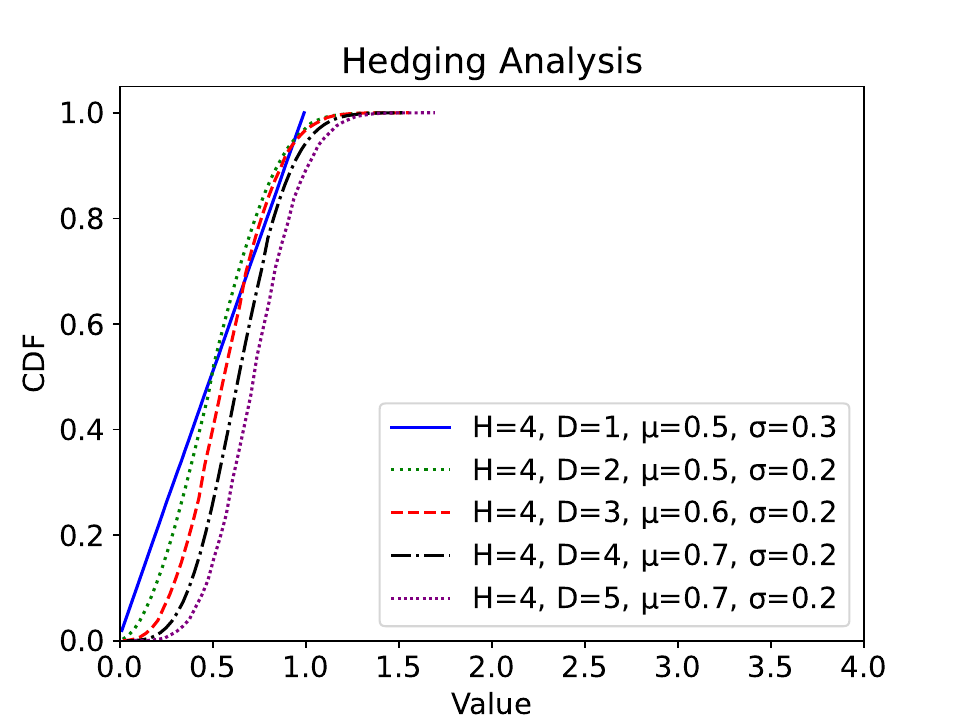}
        \caption{Hedging limits the impact of depth on the latency and its variance.}
        \label{fig:rv:h4_d}
  \end{subfigure}\hspace{0.3cm}
  \begin{subfigure}[]{0.3\textwidth}
        \includegraphics[width=\textwidth]{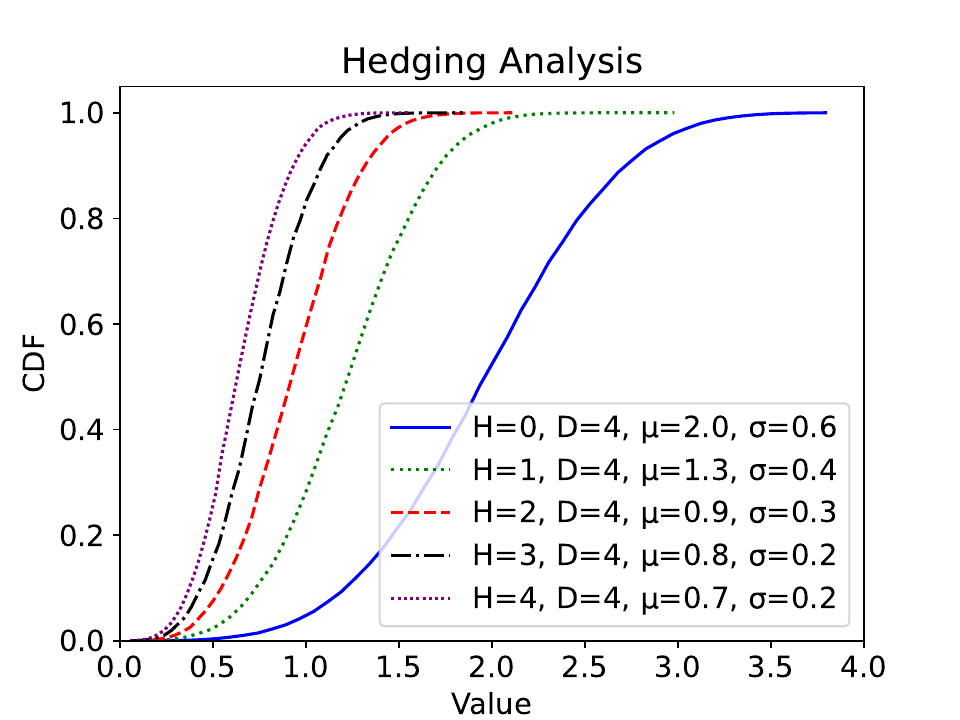}
        \caption{For a fixed depth, increasing H improves the latency and its variance.}
        \label{fig:rv:d4_h}
  \end{subfigure}
  \caption{Analyzing VM Hedging. A Monte Carlo simulation with 100k iterations was used.}
  \label{fig:monte-carlo}
\end{figure*}

When VM hedging is enabled, we may model the latency experienced by a receiver as follows.

We use function $\texttt{L}(a, b)$ to represent the latency from node $a$ to node $b$. We use $S$ to represent the root node (i.e., the sender) in Figure~\ref{fig:tree_overview}, and $P_i^j$ to represent the node $i$ (i.e, a proxy or a receiver) in Layer $j$. Then, given a node $P_i^n$, the end-to-end latency from the root sender to this node can be recursively defined as the following random variable ($\mathbb{U}$ denotes uniform random distribution): 


$$\texttt{L}(S, P_i^n) = \min_{0\leq j \leq H}\left\{ \texttt{L}(S, {P_{ (\lfloor i/F\rfloor -j) }^{n-1} })  + \texttt{L}(P_{ (\lfloor i/F\rfloor -j) }^{n-1}, P_i^n) \right\} $$

$$\forall i, j: \texttt{L}(S, P_i^{0}) \sim  \mathbb{U}  \text{ and } \texttt{L}(P_i^{n-1}, P_j^{n}) \sim \mathbb{U} $$

Each $\texttt{L}(P_i^{n-1}, P_j^{n})$ is assumed to be independent and identically distributed (IID). However, the latency is impacted by the order in which a parent proxy sends out the messages. This happens because of the replication and the transmission delay of sending messages. If the number of downstream nodes for a proxy stays small, we can ignore this order and assume $\texttt{L}(P_i^{n-1}, P_j^{n})$ to be IID. 

Achieving a low variance of $\texttt{L}(S, P_i^n)$ would mean that the latency over time may not deviate much from the expected value, helping in achieving consistent latency over time. It also shows that different VMs (at the same level of the tree) in \systemname{} may not experience latency significantly different from each other, reducing the difference between the maximum and minimum latency experienced among all the receivers for a multicast message. 

We run a Monte Carlo simulation of $\texttt{L}(S, P_i^n)$ random variable with different values of $H$ and $D$ to understand its behavior. The simulation is run for 100k iterations. Based on the simulation results (Figure~\ref{fig:monte-carlo}), we have three main takeaways.

\textbf{No Hedging $\approx$ High Latency Variance:} With no hedging, the depth of a tree and latency variance are directly correlated. Figure \ref{fig:rv:h1_d} plots the CDF for $\texttt{L}(S, P_i^n)$ and shows mean value $\mu$ and standard deviation $\sigma$ for different configurations of the multicast tree. As $D$ grows larger, we see both $\mu$ and $\sigma$ increase distinctly, indicating that just a proxy tree (i.e., no hedging) suffers from more latency variance when the tree scales up.

\textbf{Hedging $\approx$ Low Latency Variance:} Our hedging makes the correlation between $D$ and $\sigma$ become less significant. In Figure \ref{fig:rv:h4_d}, we can see the latency distribution becomes \emph{narrow} (i.e., reduced $\sigma$) as $H$ grows from 0 to higher values.

\textbf{Low Overall Latency \& Diminishing Returns on $H$:} In Figure~\ref{fig:rv:d4_h}, we fix $D$ and keep increasing $H$. Figure~\ref{fig:rv:d4_h} shows that hedging not only reduces the latency variance (i.e., the distribution becomes narrower), but it also helps to reduce the overall latency (i.e., the distribution moves leftwards). However, as $H$ grows larger, the incremented performance gains diminish, and most performance improvement is obtained when $H$ grows from 0 to 1. It shows that a small value of $H$ (>0) is enough to reap the benefits of VM hedging. This is useful because a small $H$ saves bandwidth.  



\input{sharded_throughput}

\vspace{0.5cm}
\section{Receiver Hedging}
\label{app:receiver_hedging_sync}

Each receiver VM belonging to an MP runs the same trading algorithm, receives market data, and generates orders. If any one VM belonging to an MP suffers from performance variation or data is not disseminated fairly to it, then the MP's other VM would help retain good performance, with high probability. This technique improves the performance guarantees significantly at the tail. 

Receiver hedging doubles the number of receiver VMs, which increases the dollar cost. On the other hand, due to the scaling properties of the proxy tree, the increase in OML (overall multicast latency) is negligible when the number of receiver VMs doubles. Moreover, when we consider latency for an MP as the minimum latency experienced by any of its two receiver VMs, the OML improves (\S\ref{evals_rcvr_hedging}).

As two VMs belonging to an MP participate in trading simultaneously, (i) their state (e.g., LOB snapshot used for deciding when to place orders) need to be in sync and (ii) we need a way to discard any duplicate order submissions i.e., if both VMs of an MP submit the same order, we need a way to identify and discard one copy. 

\Para{Order De-duplication: } If a receiver VM computes a hash of all market data points used to create an order and attaches the hash to the order message, the proxies can efficiently de-duplicate such orders.

\Para{State Synchronization: } Most messages originating from the exchange server that may modify the state at receiver VMs are observed by both VMs in an MP, as outbound packet losses are negligible. Moreover, since the exchange server sequences all messages, receiver VMs can easily detect rare gaps and synchronize with each other to reconcile their state.

\section{Multicast Packet Losses}
\label{app:packet_losses}

We define a packet as lost if at least one of multicast receivers do not receive it. In our 20 seconds benchmark on GCP, using \texttt{c2d-highcpu-8} instances, with 100 multicast receivers, we observe small packet losses: $\leq0.007\%$ packets lost with a multicast message rate of 350K MPS (utilizing $\approx80\%$ egress of each proxy). At smaller message rates e.g., 100K MPS, we do not observe a single packet loss. 

On AWS, using \texttt{c5.2xlarge} instances, we observe slightly higher packet losses: $0.008\%$ lost packets with a multicast rate of 10K MPS and $1.7\%$ lost packets with a rate of 100K MPS. Although we are exhausting  $\approx37\%$ egress of a proxy at 100K MPS, we suspect that the increased losses are due to the packet per second quotas implemented by AWS~\cite{github_pps_limits_ec2}. 







\input{trustmodels}

\section{Cost of Proxy Hedging} 
\label{app:cost_hedging}

Proxy hedging lowers multicast throughput in proxy VMs due to redundant tasks like sending messages to nieces. Reclaiming this lost throughput requires $H$ parallel proxy trees with a shared root (sender) and leaves (receivers), assuming receivers can handle the traffic. Table \ref{table:cost_proxies} details the cost of proxies on AWS, based on the \texttt{c5.2xlarge} instance at \$0.34/hour for 100 and 1000 multicast receivers which is minuscule compared to on-premises exchange's infrastructure cost and colocation fees~\cite{cost_of_exchange, dbo_prateesh}. 

\begin{table}[h]
    \centering
    \scriptsize
    \begin{tabular}{cccc}
        \toprule
        N & \# of proxy VMs & \# of proxy VMs & Cost/hour\\
          & (H=0) &  (H=2) & (H=2) \\
         \midrule
        \rowcolor{lightgray}
        100 & 10 & 30 & \$10.2 \\
        1000 & 110 & 330 & \$112.2 \\
        \bottomrule
    \end{tabular}
    \caption{Cost of proxy hedging}
    \label{table:cost_proxies}
\end{table}

\section{LOQ Proof's Intuition}
\label{loq_proof_intuition}

For any contiguous sequence $S$ of orders with the same mid-price (and therefore same value of $I_m$), only the critical orders will actually be executed (i.e. matched with another order) by the ME, in order of their timestamps. Consider a non-critical bid with value $b < m - w$. Since the lowest existing asking price before the sequence is executed is, by definition, greater than $m$, this bid cannot match with any existing ask. Then, since all asks in the sequence have value $a \geq m - w$, the non-critical bid cannot match with any other asks in its sequence. Non-critical orders in $S$ will only be matched with orders that have strictly greater $I_m$ while all orders with higher $I_m$ have lower priority than all orders with lower $I_m$. 

Within $S$, as long as the critical orders are processed in the sorted timestamp order, the resulting executions will be the same as executing \emph{all} orders sorted by timestamp. Furthermore, if the non-critical orders are also processed in the sorted timestamp order, the state of the order book will be the same as processing each order in $S$ sorted by timestamp. The limit order book is first sorted by value (descending for bids and ascending for asks) and then by timestamp; all the critical orders will be sorted first and then the non-critical orders will be sorted after.

The LOQ construction holds the property that the critical and non-critical orders are sorted by timestamp separately when dequeued. Then, the sequencer at each node ensures that this property is maintained when combining multiple streams consisting of LOQ outputs. Finally, since $I_m$ is always increasing with timestamp, the sequence of orders that the ME can be partitioned into contiguous sequences of orders with the same value of $I_m$. Therefore, since the executions within and the state of the order book between each partition equal the executions and state if the orders were processed sorted by timestamp, the order of executions across the entire sequence or requests will be the same. 


%% file: sharded_throughput.tex
\section{Optimizations for High Throughput}
\label{sec:thpt}
\label{app:thpt}

\Para{Decoupling DPDK Tx/Rx processing:} In the outbound direction of \systemname{}, we leverage DPDK to bypass the kernel and multicast market data in low latency. We adopt the high-performance lockless queue~\cite{dpdk-lockless-qu} provided by DPDK to decouple the Tx/Rx processing logic. On each 8-core proxy VM in \systemname{}, we allocate one core (i.e., one polling thread) for Rx and 6 cores for Tx (while 1 core is reserved for logging/monitoring processes). The Rx thread keeps polling the virtual NIC to fetch the incoming messages and dispatch each message to one Tx thread via the lockless queue, which continues to replicate and forward the messages. 

\Para{Minimizing packet replication overheads:}  When a Tx thread is replicating the message, instead of creating $F$ packets with each containing one complete copy of the message, we use a zero-copy message replication technique. For each packet, we remove the first few bytes that contain Ethernet and IP header and then invoke $rte\_pktmbuf\_clone()$ API to make several shallow copies of the packet, equal to the number of downstream nodes of a proxy. We allocate small buffers for new Ethernet and IP headers (from pre-allocated memory pools) and attach each pair of these buffers to one shallow copy created previously. Then we configure the headers (i.e., writing the appropriate destination addresses) and use $rte\_eth\_tx\_burst()$ API to send the packets out.

\Para{Parallelizing multiple multicast trees:} To further improve the throughput, we inherit the sharding idea used by CloudEx \cite{cloudex-hotos}. Since each piece of market data is associated with one trading symbol (e.g., \$MSFT, \$AAPL, \$AMD), we can employ multiple trees in parallel to multicast the market data associated with different symbols. In this way, \systemname{}'s throughput scales horizontally by adding more multicast trees.

%% file: trustmodels.tex
\section{\sysname{} Trust Models}
\label{sec:trust_model}
In an ideal trust model, an exchange wouldn’t need to rely on MPs to adhere to protocols like accurately time-stamping trade orders or withholding market data processing until a set deadline to ensure fairness. Likewise, MPs wouldn't have to disclose their trading algorithms to the exchange. These guarantees should be achieved without incurring any performance overhead, such as increased latency or reduced throughput. 

\begin{figure}[t]
        \centering
        \vspace{-0.2cm}
        \includegraphics[width=0.9\linewidth]{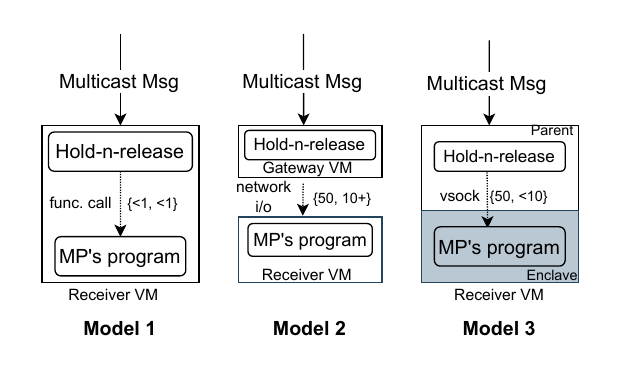}
        \vspace{-0.5cm}
        \caption{Architectures corresponding to each trust model. Dotted arrows represent messages going from an exchange's programs to an MP's program where the type of message is represented on the left side and \{$a$, $b$\} denotes $a$ \SI{}{\micro\second} of latency and $b$ \SI{}{\micro\second} of jitter for messages.}
        \label{fig:model_archs}
        \vspace{-0.3cm}
\end{figure}

An ideal model should also not introduce any \textit{jitter} for packets going from the hold-and-release (the exchange's program to delay release of market data) to MPs' trading algorithms. Fairness is achieved at the level of the hold-and-release program using CloudEx's time synchronization mechanisms, which \sysname also leverages. To ensure fairness at the level of trading algorithms, there should be a constant latency between the exchange program and the trading algorithm. Achieving this ideal model is impractical due to the tension between security and performance that we present. We propose three trust models (Fig.~\ref{fig:model_archs}), discuss their respective trade-offs, and offer our recommendation. 

\Para{Model 1: MPs give their programs to the exchange.} Exchange controls the VMs where the MPs' trading programs run. The exchange runs hold-and-release in these VMs and then forwards the multicast messages to the MPs' programs. No significant latency or throughput overhead is incurred. The jitter between the exchange's program and the MP's program can be minimized to be non-significant ($\leq$\SI{1}{\micro\second}) as forwarding messages between the two programs can be mere function calls. However, MPs have to reveal their proprietary trading programs to the exchange in this model. 

\Para{Model 2: Separate gateways for hold-and-release.}  Model 2 deploys gateways between the exchange's infrastructure (ME and proxy tree) and the VMs where MPs' algorithms run. The exchange controls the gateways hosting the hold-and-release programs, while MPs control the VMs running their algorithms. It avoids the need for MPs to reveal their programs or for the exchange to trust MPs with hold-and-release. However, it introduces latency between the hold-and-release program and MPs' programs, as they run in separate VMs. Throughput is unaffected, depending on VM bandwidth, but OWD between VMs is around \SI{50}{\micro\second}, with high jitter due to cloud latency fluctuations and spikes. In the absence of the spikes, jitter may be in the tens of microseconds. Even with simultaneous delivery at the gateway level, fairness at the receiver VM level is not guaranteed due to this jitter.

\Para{Model 3: Trading programs run in secure enclaves} Model 3 leverages the confidential computing capabilities of VMs equipped with secure enclaves such as AWS Nitro Enclave. An enclave can only talk to its associated (parent) VM (and AWS Nitro Hypervisor). The exchange owns the VMs subjecting incoming messages to hold-and-release before forwarding them to the respective enclaves within the VMs. The MPs' program executes within the enclave.\footnote{Loaded via remote attestation mediated by AWS Nitro Hypervisor} MPs do not have to reveal their trading programs to the exchange and the exchange does not have to rely on the MPs to run hold-and-release. Although perfect security boundaries are achieved, performance overhead is incurred: high latency, extremely low throughput, and non-negligible jitter. 

The latency between a VM and an enclave is similar to the latency between two VMs, making it comparable to Model 2. However, Model 3 benefits from reduced jitter between a VM and its enclave, likely because communication between them does not traverse the network. Optimizations described next further reduce the jitter. Nonetheless, throughput between a parent VM and its enclave is significantly lower (about 70\% lower) compared to the parent VM's ingress, as noted in \cite{nitriding}.

\begin{figure}[t]
    \centering
    \includegraphics[width=0.7\linewidth]{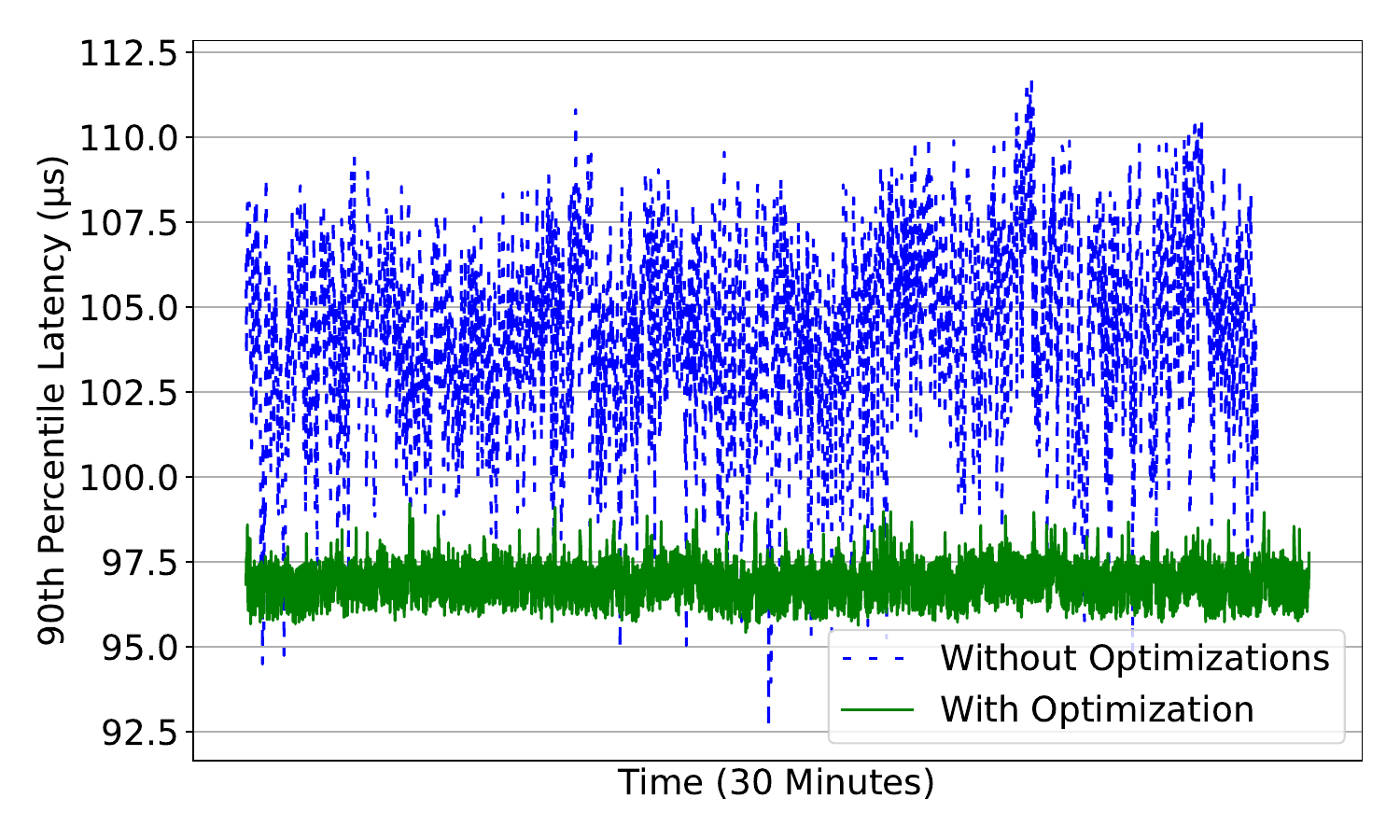}
    \vspace{-0.5cm}
    \caption{RTT between a VM and its enclave stabilizes}
    \label{fig:windowed_enclave_latency}
    \vspace{-0.5cm}
\end{figure}

\Para{Reducing jitter between a VM and an enclave} We reduce the latency variance between a VM and its associated enclave with some optimizations: isolating CPUs, reducing scheduling-clock ticks, and pinning threads to cores. Figure \ref{fig:windowed_enclave_latency} shows the 90th percentile latency between a parent VM and its associated enclave for each tumbling window of \SI{1}{\second}. After the optimizations, the latency becomes much more predictable. We observe a jitter (the difference between $90p$ and $50p$ latency) of $\leq\SI{10}{\micro\second}$. 


\Para{Recommended Model.} Optimal performance is attained with Model 1, though MPs need to reveal their programs to the exchange. If an exchange is obliged to not reveal its clients information to third parties (potentially binding due to Section 6801 of 15 U.S. Code in some territories~\cite{sec6801}), this model should be adopted in practice. We use Model 1.